 \documentclass[final,3p,times]{elsarticle}

\usepackage{lineno,hyperref}
\usepackage{amsmath}
\usepackage{amsfonts}
\usepackage{amssymb}
\usepackage{graphicx}
\usepackage{float}    
\usepackage{enumitem} 
\usepackage{nameref}
\usepackage{mathtools}
\usepackage{verbatim}
\usepackage{caption}
\usepackage{subcaption}
\usepackage{mathrsfs}
\usepackage{multicol}
\usepackage[table]{xcolor}
\usepackage{multicol}

\journal{arXiv}

\begin{document}

\begin{frontmatter}

\title{Variable-Pitch Power Regulation of Tethered-Wing Systems Based on Robust Gain-Scheduling H-infinity Control}
\author{Mani Kakavand} %
\ead{manikakavand@gmail.com}
\author{Amin Nikoobin\corref{mycorrespondingauthor}}
\cortext[mycorrespondingauthor]{Corresponding author} 
\ead{anikoobin@semnan.ac.ir}
\address{Faculty of Mechanical Engineering, Semnan University, Across Sookan Park, Semnan, Iran, 3513119111}
\begin{abstract}
In this paper, we deal with the power regulation of tethered-wing systems and demonstrate advantages of variable-pitch control in mitigating the dynamic mechanical loads and power fluctuations.
The proposed scheme is based on a strategy that maximizes the energy capture during low-speed wind and prevents overloads during the high-speed wind.
To realize this strategy, we use a tether reeling-speed controller to track the optimal generator speed during low-speed wind and a MIMO speed-force controller for power limitation during high-speed wind.
The controllers are synthesized using $ \mathcal{H}_\infty $ method and are based on a linear parameter varying (LPV) system that expresses the flexible dynamics of system as a function of the tether's length and force.
Using this method, the controllers are made robust with respect to dynamic and parametric uncertainties and the wing's pitch angle activity is minimized during high-speed wind.
We carry out extensive simulations to demonstrate the controllers' performance.
These include the implementation of the scheme in a detailed 3-dimensional tethered-wing system simulator with a realistic turbulent wind field.
\end{abstract}




\end{frontmatter}




\section{Introduction}
\begin{figure}[h!]
	\centering
	\includegraphics[width=0.5\linewidth]{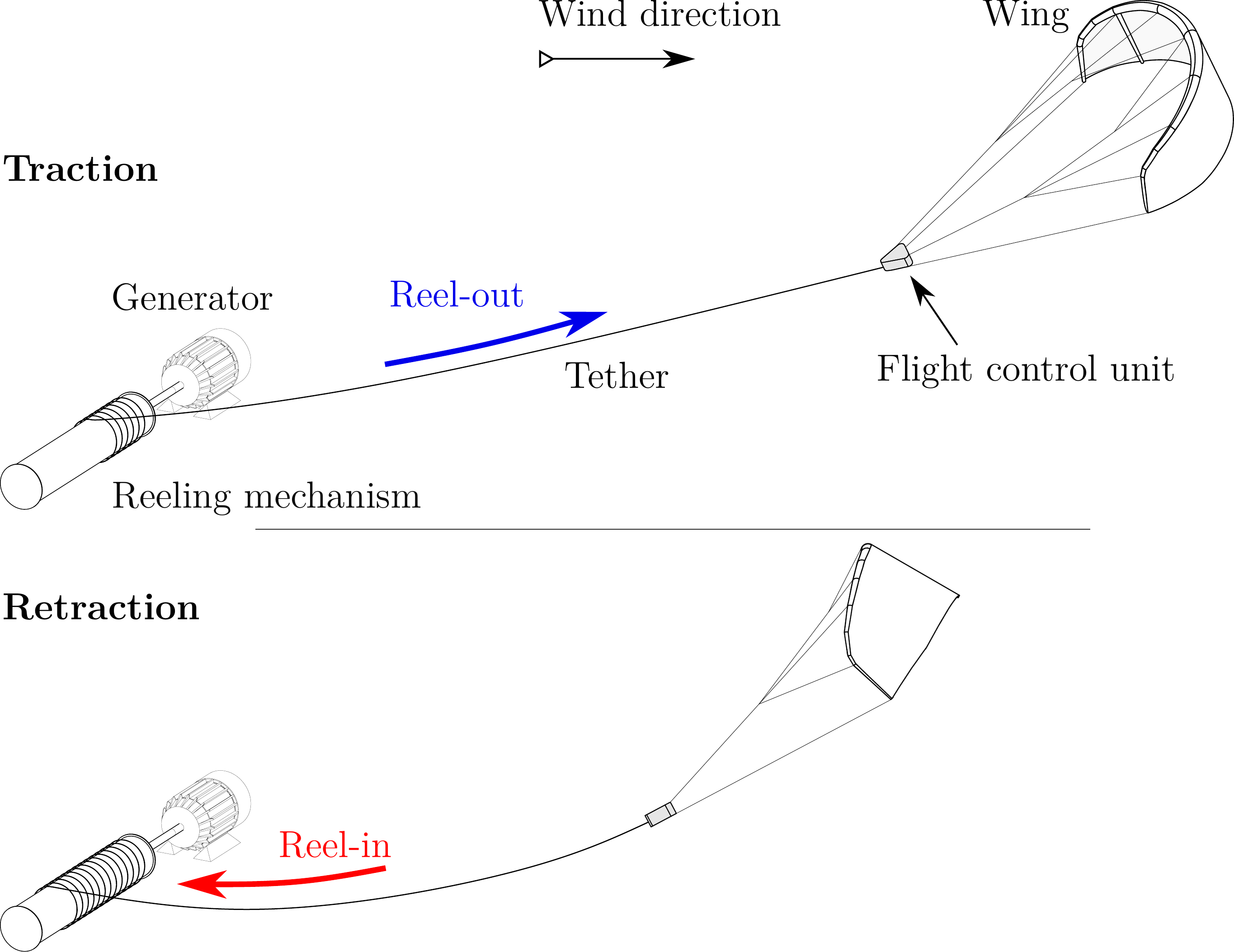}
	\caption{The cyclic operation of a kite-generator featuring a soft kite}
	\label{fig:concept}
\end{figure}

Airborne wind energy (AWE) is a developing technology with the aim of harnessing the strong and steady winds of high altitudes that are inaccessible by the conventional wind turbines due to their structural constraints \cite{Fagiano_2010}.
Deploying a wing via a long tether to the high altitudes ($ 200-800\ m $ from the ground) to capture the wind energy is the solution provided by the ground-based generation tethered-wing system (aka kite-generator).
The idea is to transmit the wing's aerodynamic force by the tether to a reeling-mechanism that is coupled to a generator.
The system has to operate by repeating a traction-retraction cycle as shown in Fig.~\ref{fig:concept}.
During the traction phase, wind energy is converted into electrical energy as the wing flies in a cross-wind motion and the tether is being reeled out.
During retraction, electrical energy is consumed to reel in the tether as the wing flies to the starting point of the course.
This 2-phased cycle is planned so that the net generated energy is positive.
See reference \cite{Schmehl_2013} for an in-depth treatment of the operating principle of kite-generators and cross-wind flight and reference \cite{cherubini2015airborne} for a review of AWEs.

The operation of kite-generators depends on the control systems that guide the wing along the prescribed path and regulate the amount of the generated power.
The former can be performed by manipulating the wing's control surfaces (or the bridle lines in the case of soft wings) and the latter is usually done by adjusting the generator's torque or speed.
The common method to tackle the control problem is the use of modular control architectures, in which separate controllers are employed for the wing and the generator in addition to the high-level controllers that supervise the entire operation.
A survey of tethered-wing power systems' control can be found in reference \cite{vermillion2021electricity}. 

In the literature of tethered-wing systems, the power regulation problem (aka winch control) has been addressed by solutions having various degrees of sophistication.
In many cases, the solutions are presented without addressing the dynamic behavior of the controlled system.
This is while the reeling-mechanism and the tether comprise a flexible system whose dynamics include vibrational modes that can be excited during the operation.
Such vibrations have indeed been reported in the experiments of references \cite{oehler2019aerodynamic} and \cite{malz2019reference}.
Another observation that illustrates the importance of considering the dynamics of the controlled system is the existence of a trade-off between the smoothness of the tether's force and the reeling speed, i.e., a smooth speed tracking causes large variations in the force and vice versa \cite{rapp2019cascaded}.
Without a description of the dynamics of these two variables tuning the controllers has to be performed by trial and error.
The dependency of the system's dynamics on its operating point, as will be discussed in this paper, further complicates the problem.

An instance of control design without directly considering the dynamics of the controlled system is found in \cite{erhard2015flight}, in which the on-line computation of the optimal reeling-speed and a simple mechanism to prevent force overloads are discussed.
This has led to smooth reeling-speed tracking but severe fluctuations in the tether's force and consequently the generated power.
A similar approach for reeling-speed control is used in \cite{fechner2015dynamic} with the use of a parameter-varying PID controller.
Employing a tether's force controller for the traction phase is a common approach as proposed in \cite{todeschini2021control}.
The reference has used a reeling-speed controller for the retraction phase.
The application of model predictive control to kite-generators, such as \cite{Lago_2018}, also fall within the same category as they generate reference signals for the low-level controllers (speed/force controllers) without considering the dynamics of the controlled system.

An example of model-based power regulation can be found in reference \cite{fechner2016methodology}, in which the aerodynamics of the wing is considered and different controllers are used for high and low tether force.
Reference \cite{rapp2019cascaded} has employed a tether force controller for the traction phase and a model-based feed-forward speed controller for retraction in which the tether's force is treated as a measurable disturbance.
An instance of active control of the angle of attack for rigid wing tethered-wing systems can be found in the same reference.
A PID force controller with a feed-forward from a reel-out speed predictor is proposed in \cite{fechner2014feed}.
Having that said, none of the model-based control designs address the issue of longitudinal oscillations or provide a study of such mode shapes that may get excited during the system's operation.
Additionally, all of the mentioned references have used the generator's torque as the only control input to regulate the power and none has proposed a method based on the manipulation of the angle of attack during the traction phase.
This is in contrast with the contribution of this paper, in which we propose a power regulation scheme that utilizes the wing's pitch angle in addition to the generator's torque.
The only instance of active pitch control of tethered-wing systems appears in reference \cite{buffoni2014active}.
The authors have proposed a pitch adaptation algorithm that maximizes the tether force by incrementally changing the pitch angle until the optimum value is reached.

Another significant difference between our approach and the works of literature is our reliance on a dynamic model that expresses the longitudinal dynamics of the system for control design.
With that purpose, we formulate the system's non-linear behavior as a linear parameter varying system (LPV).
This particular formulation allows for the application of various control methods.
For instance, see the the application of shifting state-feedback control or model predictive-based gain-scheduling control of LPV systems in \cite{Ruiz_2022} and \cite{Vafamand_2018}, respectively. 

After providing a discussion of the open-loop characteristics of the LPV model, we express the control objectives as an LMI optimization problem using the $ \mathcal{H}_\infty $ method.
This is a powerful method to obtain controllers with the ability to minimize the effects of disturbances and uncertainties \cite{Sedhom_2020} and with a wide range of applications.
For instance, see the application of $ \mathcal{H}_\infty $ method to robust control of an induction generator \cite{Demirtas_2021}, multi-motor servomechanisms \cite{Wang_2018}, power systems \cite{Chang_2020}, and adaptive control of the internet network \cite{Liu_2018}.
Another distinctive feature of this method is the applicability of the long-established tools of linear control theory.

Using the $ \mathcal{H}_\infty $ method, we synthesize two gain-scheduled output-feedback controllers for the the operation in the low- and high-speed winds; a SISO reeling-speed controller for the below-rated operation, and a multivariable reeling-speed and tether force controller for the above-rated condition.
In the case of reeling-speed control, the LMI problem is formulated to make a trade-off between the reference tracking performance and the fast variations in the tether's force.
In the multivariable case, in which there is no a priori allocation of the control inputs to the controlled variables, the aim is to achieve satisfactory tracking performance while reducing the wing's pitch angle activity.
Using the small-gain theorem, the controllers are synthesized so that they are robust with respect to the unmodeled dynamics and the parameter uncertainties.
The proposed method is more practical for the systems with a rigid-wing, due to the difficulty of measurement and control of the angle of attack for the soft kites \cite{buffoni2014active}.
See \cite{rapp2019cascaded} for instance of active control of the angle of attack for rigid wing tethered-wing systems.


In summary, the main contributions of this paper are:
\begin{itemize}
	\item the formulation and characterization of the flexible longitudinal dynamics of the tethered-wing systems;
	\item proposing a variable-pitch power regulation for tethered-wing power systems;
	\item illustrating the advantages of active pitch control for reducing the dynamic mechanical loads and power fluctuations;
	\item providing a method for the systematic treatment of the trade-offs in control design for the power regulation problem.
\end{itemize}

The outline of this manuscript is as follows. 
In section \ref{sec2}, we develop the dynamic longitudinal model.
The open-loop characteristics of this model are discussed in \ref{sec3}.
A quick review of the $ \mathcal{H}_\infty $ gain-scheduling control method along with the explanation of the higher-order singular value decomposition method used to transform the LPV system into an affine polytopic one is presented in \ref{gss}.
Section \ref{cs} is dedicated to an explanation of the power regulation strategy. 
The application of the control method to the affine polytopic representation of the dynamics of the system is provided in section \ref{sec6}. 
And finally, the simulation results and a short conclusion section are given in section \ref{sec7} and \ref{sec8}, respectively.
The constants, parameters, and the set of LMIs used for controller synthesis are provided in the \nameref{Appendix}.
\section{Longitudinal Dynamics Modeling}\label{sec2}
In this section, we develop a model to capture the dynamics of the system that is comprised of wing and the tether, and the reeling-mechanism. 
Two simplifications are considered for that purpose;
Firstly, we neglect the motion of the wing and the tether in the perpendicular direction to the tether and refer to the remaining dynamics as longitudinal dynamics.
This assumption is made so that we can get rid of any dynamics that cannot be seen from the point of view of the generator.
Considering that, the motion of the wing perpendicular to the tether is irrelevant to the problem at hand.
Note that this simplification also leaves out the tether's transverse waves (vibrating string modes).
Since these are much slower than the longitudinal waves \cite{S_nchez_Arriaga_2019}, we do not expect them to affect the controller's performance.
The next simplification is that we express the aerodynamic force directly as a function of the angle of attack and the wind speed and neglect any intermediary dynamics.
We will deal with this assumption by making the controllers robust with respect to the ensuing errors.
This subject will be discussed in sections \ref{sec6} and \ref{sec7}.

\begin{figure}[h]
	\centering
	\includegraphics[width=0.55\linewidth]{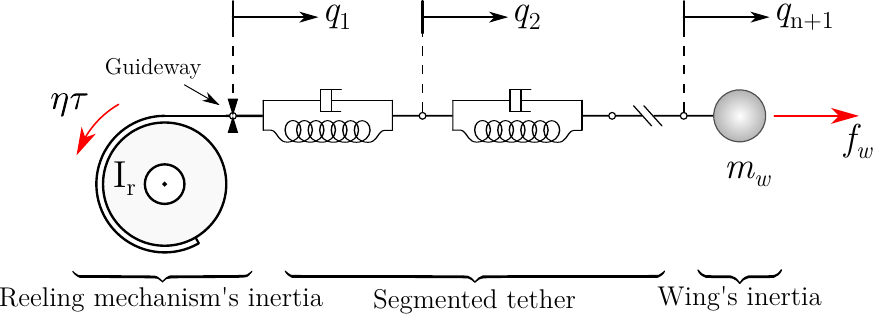}
	\caption{Schematic of the longitudinal dynamic model and the external forces}
	\label{fig:tethered-wing}
\end{figure}

Figure \ref{fig:tethered-wing} illustrates a schematic of the longitudinal model, which consists of:
\begin{enumerate}[label=(\roman*)]
	\item the reeling-mechanism represented by the moment of inertia $ I_r $;
	\item the tether as a series of $ n $ spring-damper segments with unstretched length $ \ell $ and linear density $ \mu $ that define the generalized coordinates $ q_i\in\mathbb{R} $ for $ i=1,\ \dots n+1 $, henceforth referred to as nodes; 
	\item the point-mass $ m_w $ with the coordinate $ q_{n+1} $ that represents the motion of the wing in the direction of the tether.
\end{enumerate}

The external forces acting on the system are the generator's torque $ \tau $ applied to the reeling-drum through a gearbox with ratio $ \eta $ and the aerodynamic force magnitude $ f_{w} $, as a function of wind speed and the angle of attack that is exerted on the wing with mass $ m_w $.
To derive the equations of motion, we use the Lagrange method as follows.
\subsection{Lagrangian}
The contribution of each component of the dynamic model to the Lagrangian, considering the schematic of Fig.~\ref{fig:tethered-wing}, is as follows.
\paragraph{Reeling Mechanism}
It is assumed that the undeployed portion of the tether is inelastic and wound around the drum with radius $ r $.
Given that the velocity of the tether's first node is $ \dot{q}_1 $, the drum would have the angular velocity $ \dot{q}_1/r $ and therefore the reeling-mechanism's kinetic energy is given by
\begin{equation}\label{rm_kinetic}
	T_r = \dfrac{1}{2}I_r(\dfrac{\dot{q}_1}{r})^2.
\end{equation}
Note that the reeling-mechanism's moment of inertia varies as a function of the length of the tether that is wound around the drum.
That is, $ I_r = I_d +\mu(L_t-L_d)r^2 $, in which $ I_d $ is the moment of inertia of the rotating parts sans the wound tether, $ L_t $ is the total length of the tether, and $ L_d = n\ell $ is the length of the deployed portion of the tether.
\paragraph{Tether} The position of the material points on the $i$-th tether segment with respect to the stationary guideway is given by
\begin{equation}\label{t_position}
	p_i(\zeta)=(1-\dfrac{\zeta}{\ell})q_i + (\dfrac{\zeta}{\ell}) q_{i+1},
\end{equation}
for $ \zeta\in[0\ \ell] $, where $ q_i\in\mathbb{R} $ denotes the position of the $i$-th node.
Using this expression, we derive the velocity of each material point on the tether as
\begin{equation}\label{t_velocity}
	\dot{p}_i(\zeta)=(1-\dfrac{\zeta}{\ell})\dot{q}_i + (\dfrac{\zeta}{\ell}) \dot{q}_{i+1},
\end{equation}
in which we have neglected the time-dependent variations of $ \ell $  to simplify the derivation.
The kinetic energy of the deployed portion of the tether is given by the sum of the kinetic energy of the $ n $ segments as
\begin{equation}\label{t_kinetic}
	T_t = \sum_{i=1}^{n}\int_{0}^{\ell}\dfrac{1}{2}\mu \dot{p}_i(\zeta)^2 d\zeta .
\end{equation}

To find the potential energy, we calculate the strain of the tether $i$-th segment as
\begin{equation}\label{t_strain}
	\epsilon_i(\zeta) =\lim_{\Delta \zeta \rightarrow 0} \dfrac{p_i(\zeta+\Delta\zeta)-p_i(\zeta)}{\Delta \zeta} -1 = \dfrac{\partial p_i(\zeta)}{\partial\zeta}-1,
\end{equation}
using which we have
\begin{equation}\label{t_potential}
	V_t = \sum_{i=1}^{n}\int_{0}^{\ell}\dfrac{1}{2}E_tA_t\epsilon_i(\zeta)^2 d\zeta,
\end{equation}
where $ E_t $ is the tether's Young modulus and $ A_t $ is its cross-section area.
\paragraph{Wing} Given that the point-mass $ m_w $ is attached to the $ (n+1) $-th node, the wing's kinetic energy for the longitudinal motion is
\begin{equation}\label{w_kinetic}
	T_w = \dfrac{1}{2}m_w\dot{q}_{n+1}^2.
\end{equation}
Finally, using the Eqs.~\eqref{rm_kinetic}, \eqref{t_kinetic}, \eqref{t_potential}, and \eqref{w_kinetic}, we express the system's Lagrangian as 
\begin{equation}\label{lagrangian}
	L_{tot} = T_r + (T_t - V_t) + T_w.
\end{equation}
\subsection{Aerodynamics}
\begin{figure}[h]
	\centering
	\includegraphics[width=0.45\linewidth]{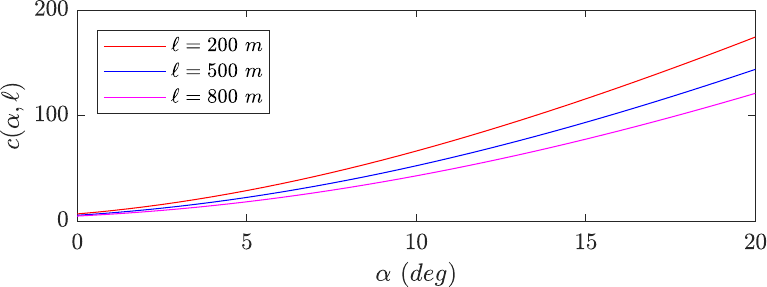}
	\caption{The aerodynamic coefficient $ c(\alpha,\ell) $}
	\label{fig:00-aerodynamicc}
\end{figure}

The aerodynamics of the wing is specified by its lift and drag coefficients as functions of the angle of attack $ \alpha $.
We use the typical aerodynamic coefficients 
\begin{equation}\label{lift_drag}
	C_L(\alpha) = (4.1\times10^{-2})\alpha + 0.2,\quad C_D(\alpha) =(1.74\times10^{-3})\alpha+0.29 ,
\end{equation}
for the range $ \alpha \in [0, 20] $ degrees to calculate the wing's aerodynamic force.
It is known that the contribution of the tether's drag is equivalent to a quarter of the tether's length moving with the same speed as the wing \cite{van2019quasi}.
Including this effect in $ C_d $, we find the airborne system's total aerodynamic drag as
\begin{equation}\label{Cdeq}
	C_{D,eq}(\alpha,\ell) = C_D + \frac{d_t L_d}{4S_w}C_{D,t},
\end{equation}
where $ d_t $ is the tether's diameter, $ S_w $ is the wing's effective area, and $ C_{D,t} $ is tether's drag coefficient.
Hence, the magnitude of the aerodynamic force is given by
\begin{equation}\label{aero}
	f_{w} = \dfrac{1}{2}\rho S_w\sqrt{C_L(\alpha)^2+C_{D,eq}(\alpha,\ell)^2} v_{app}^2,
\end{equation}
in which $ v_{app} $ is the wing's apparent wind speed that can be expressed as \cite{van2019quasi}
\begin{equation}\label{v_app}
	v_{app} = \sqrt{1+E_{eq}(\alpha,\ell)^2}(v_{w}\cos\phi-v_{r}),
\end{equation}
where $ v_{k} $ is the speed of the wing in the tether's direction, $ E_{eq}(\alpha,\ell)=C_L(\alpha)/C_{D,eq}(\alpha,\ell) $ is the wing's aerodynamic efficiency, and $ \phi $ is the elevation angle. 
Note that we have dropped the dependency on the wing's azimuth angle from the original form of Eq.~\eqref{v_app} for simplicity.
Finally, replacing $ v_{r} $ with the velocity of the tether's last node, we express the aerodynamic force as
\begin{equation}\label{f_w}
	f_{w}(v_{w,t},\alpha,\ell,\dot{q}_{n+1})= c(\alpha,\ell)(v_{w,t}-\dot{q}_{n+1})^2,
\end{equation}
where $ v_{w,t}=v_{w}\cos\phi $ and
\begin{equation}\label{c_aero}
	c(\alpha,\ell) = \dfrac{1}{2}\rho S_w(1+E_{eq}(\alpha,\ell)^2)\sqrt{C_L(\alpha)^2+C_{D,eq}(\alpha,\ell)^2},
\end{equation}
and $ \rho = 1.22\ kg/m^3$ is the density of air. The variation of $ c(\alpha,\ell) $ with $ \ell $ and $ \alpha $ is shown in Fig.~\ref{fig:00-aerodynamicc}.
\subsection{Equations of Motion}
The equation of motion for the generalized coordinate $ q_i $ is derived from
\begin{equation}\label{t_lagrange}
	\dfrac{\partial}{\partial t}\dfrac{dL}{d\dot{q}_i}-\dfrac{dL}{dq_i} = f_i,
\end{equation}
where $f_i$ is the $i$-th generalized force.
For nodes $ i=2,\dots,n $, this expresses the tether's structural damping that we model as a viscous damping with the coefficient $ b_t(\ell) = b_{0,t}/\ell $, i.e.,
\begin{equation}\label{t_damping}
	f_i = b_t(\ell)(\dot{q}_{i+1}-\dot{q}_i) +b_t(\ell)(\dot{q}_{i-1}-\dot{q}_i).
\end{equation}
The structural damping is included for the sake of generality and is set to zero for control design and simulation in this paper.
In the case of the reeling-mechanism, the generalized force is 
\begin{equation}\label{r_force}
	f_1 = -\dfrac{\eta}{r}\tau - b_r\dot{q}_1 -b_t(\ell)(\dot{q}_1-\dot{q}_2),
\end{equation}
where $ \frac{\eta}{r}\tau $ expresses the generator's torque and $ b_r $ is the reeling-mechanism's viscous damping coefficient.
For the last tether's node, we have
\begin{equation}\label{t_dampingn1}
	f_{n+1} =  b_t(\ell)(\dot{q}_{n-1}-\dot{q}_n) + f_{w}(v_{w,t},\alpha,\ell,\dot{q}_{n+1}).
\end{equation}

Inserting the system's Lagrangian given by Eq.~\eqref{lagrangian} into Eq.~\eqref{t_lagrange} yields $ (n+1) $ non-linear ordinary differential equations in the matrix form
\begin{equation}\label{eom}
	{M}(\ell)\ddot{{q}}+
	{C}(\ell) \dot{{q}}+
	{K}(\ell) {q}=f(\tau,v_{w,t},\alpha,\ell,\dot{q}),
\end{equation}
where $ q =[q_1\ \dots q_{n+1}]^T\in\mathbb{R}^{n+1}$ is the vector of the generalized coordinates, $ {M}(\ell) $, $ {C}(\ell) $, and $ {K}(\ell) $ are mass, damping, and stiffness matrices, respectively, and $ f = [f_1\ \dots f_{n+1}]^T$ is the vector of external forces.
\subsection{Stabilizability and Linearization}\label{ssrep}
Before linearizing Eq.~\eqref{eom} and expressing it in state-space form, we choose the state-vector such that the resulting linear system is stabilizable for the intended control schemes, i.e., the tether's force and reeling-speed control.
Specifically, the state-vector of the stabilizable representation should not include $ q $ explicitly, as it tends to infinity with constant reeling-speed.
With that intention, Eq.~\eqref{eom} is rewritten using the tether segments’ strain vector $ \epsilon=[\epsilon_1\ \dots\epsilon_n]^T $, where $ \epsilon_i=\frac{1}{\ell}(q_{i+1}-q_i) $ is defined by Eq.~\eqref{t_strain}.
This reformulation results in the non-linear equations of motion in the form
\begin{equation}\label{eom_e}
	M(\ell)\ddot{{q}}+C(\ell)\dot{{q}}+K_\epsilon{\epsilon} = f({\tau},{v}_{w,t},{\alpha},\ell,\dot{{q}}).
\end{equation}
To find a linear system that describes the local behavior of the nonlinear system around its equilibrium points, we only need to linearize the aerodynamic force $ f_w $ included in $ f_{n+1} $, since the rest of the system is already linear in terms of $ \ddot{q} $, $\dot{q}$, and $ \epsilon $.
Using bars $ (\bar{\ }) $ and hats $ (\hat{\ }) $ to denote the steady-state values and variations with respect to them, the variation of $ f_w $ is
\begin{equation}\label{hatf_w}
	\begin{split}
		\hat{f}_{w} =& \dfrac{\partial f_{w}}{\partial\alpha}\bigg |_{op}\hat{\alpha}+
		\dfrac{\partial f_{w}}{\partial v_{w,t}}\bigg |_{op}\hat{v}_{w,t}+
		\dfrac{\partial f_{w}}{\partial\dot{{q}}_{n+1}}\bigg |_{op}\dot{\hat{q}}_{n+1}\\
		=&  2c(\bar{\alpha},\ell)(\bar{v}_{w,t}-\dot{\bar{q}}_{n+1})(\hat{v}_{w,t}-\dot{\hat{q}}_{n+1})+c_\alpha(\bar{\alpha},\ell)(\bar{v}_{w,t}-\dot{\bar{q}}_{n+1})^2\hat{\alpha},
	\end{split}
\end{equation}
where $ c_\alpha({\alpha},\ell) = {\partial c({\alpha},\ell)}/{\partial\alpha} $ and $ op=(\bar{v}_{w,t},\bar{\alpha},\dot{\bar{q}}_{n+1}) $.
If we define the tether's force to be $ f_t=EA\epsilon_1 $ and consider the fact that the its steady-state value equals that of the aerodynamic force $ \bar{f}_w= c(\bar{\alpha},\ell)(\bar{v}_{w,t}-\dot{\bar{q}}_{n+1})^2 $, we can rewrite the second line of Eq.~\eqref{hatf_w} as
\begin{equation}\label{hatf_w2}
	\hat{f}_{w} =2\sqrt{c(\bar{\alpha},\ell)\bar{f}_t}(\hat{v}_{w,t}-\dot{\hat{q}}_{n+1})+\dfrac{c_\alpha(\bar{\alpha},\ell)}{c(\bar{\alpha},\ell)}\bar{f}_t\hat{\alpha}.
\end{equation}
Using Eqs.~\eqref{eom} and \eqref{hatf_w2}, we derive the linear system as
\begin{equation}\label{eq_lin}
	M(\ell)\ddot{\hat{q}}+C(\ell)\dot{\hat{q}}+K_\epsilon\hat{\epsilon} = -C_f(\bar{\alpha},\ell,\bar{f}_t)\dot{\hat{q}}+B_w(\bar{\alpha},\ell,\bar{f}_t)\hat{v}_{w,t}+B_\alpha(\bar{\alpha},\ell,\bar{f}_t)\hat{\alpha}+B_\tau\hat{\tau},
\end{equation}
where $ C_f=-\partial f/\partial \dot{q} $, $ B_w=\partial f/\partial v_{w,t} $, $ B_\tau=\partial f/\partial \tau $, and $ B_\alpha=\partial f/\partial \alpha $, in which the dependency of the functions on their variables is dropped for brevity.
If we consider the reeling-speed $ v_r=\dot{q}_1 $ and the tether's force as the measurable outputs, then the linear system's output vector is $ \hat{y}=[\hat{v}_r\ \hat{f}_t]^T $.
Next, by the definition of the state-space vector $ \hat{x}=[\hat{\epsilon}_1\ \dots\ \hat{\epsilon}_n\ \dot{\hat{q}}_1\ \dots\ \dot{\hat{q}}_{n+1}]^T $ and using the relation $ \dot{\epsilon}_i=(\dot{q}_{i+1}-\dot{q}_{i})/\ell $, the state-space representation of the linear system is 
\begin{equation}\label{state_space}
	\begin{cases}
		\begin{split}
			\dot{\hat{x}}=\left[\begin{array}{cc}
				0_{n\times n} & {L}\\
				-{M}^{-1}{K}_\epsilon & -{M}^{-1}{(C+C_f)}
			\end{array} \right]\hat{x}+&
			\left[\begin{array}{c}
				0_{n\times 1} \\{M}^{-1}{B}_w
			\end{array} \right] \hat{v}_{w,t}\\+&
			\left[\begin{array}{cc}
				0_{n\times 1} & 0_{n\times 1}\\
				{M}^{-1}{B}_\tau & {M}^{-1}{B}_\alpha
			\end{array} \right] \hat{u}, 
		\end{split}\\
		\hat{y}= C_m\hat{x}
	\end{cases}
\end{equation}
in which $ \hat{u} = [\hat{\tau}\ \hat{\alpha}]^T$, $ C_m=\partial \hat{y}/\partial\hat{x} $, and the dependency of $ M, C, C_f, L, B_w$ and $ B_\alpha $ are dropped for brevity.
The matrices used in this state-space realization are provided in \nameref{Appendix}.
Equation~\eqref{state_space} describes a linear multiple-input and multiple-output (MIMO) linear parameter varying (LPV) system whose coefficient matrices are functions of time-varying parameter set $ (\bar{f}_t, \bar{\alpha}, \ell) $, which can be regarded as the operating point of the tethered-wing system. 
\section{Open-Loop Characteristics}\label{sec3}
To investigate the open-loop behavior of the system, we construct the MIMO LPV system of Eq.~\eqref{state_space} for the dynamic model with five nodes $ (n=5) $ and study its input-output relation in the frequency domain at various operating points specified by fixed values of $ (\bar{f}_t, \bar{\alpha}, \ell) $.
The input-output relation of the LPV system in the complex frequency domain is described by 
\begin{equation}\label{tf_ol}
	\begin{split}
		\left[\begin{array}{c}
			\hat{V}_r(s)\\\hat{F}_t(s)
		\end{array} \right] 
		= \left[ \begin{array}{cc}
			G_{11}(\theta,s)&G_{12}(\theta,s) \\
			G_{21}(\theta,s) &G_{22}(\theta,s)
		\end{array}\right]\left[\begin{array}{c}
			\hat{V}_{w,t}(s)\\\hat{U}(s)
		\end{array} \right],
	\end{split}
\end{equation}
where $ s $ denotes the Laplace variable and the $ G_i(\theta,s) $s are transfer functions or matrices of appropriate size.

\begin{figure}[h]
	\centering
	\includegraphics[width=.55\linewidth]{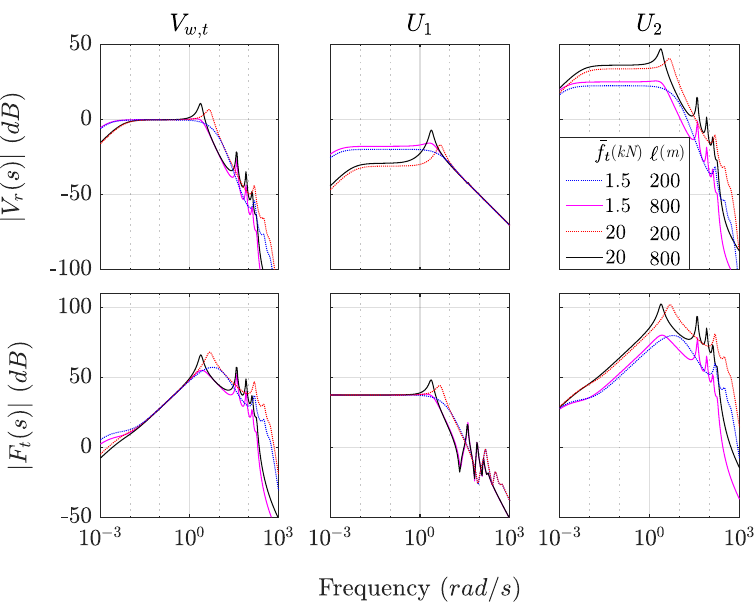}
	\caption{Frequency response of the open-loop system}
	\label{fig:olbode}
\end{figure}

Figure~\ref{fig:olbode} illustrates the magnitude of each component of the input-output transfer matrix Eq.~\eqref{tf_ol}, in which $ U_1 $ and $ U_2 $ denote the control input vector's arrays.
We have generated the results at four extremities of the operational space $ \bar{f}_t \in[1.5\ 20]\ kN$, $ \ell \in[200\ 800]\ m $, $ \bar{\alpha}= 12 \deg $. 
The system's first vibrational mode appears as a peak in all of the six plots and is characterized by an out-of-phase motion of the drum and the tether with frequency
\begin{equation}\label{wn}
	w_n\approx r\sqrt{\dfrac{E_tA_t}{L_dI_r}}.
\end{equation}
Increasing the tether force or reducing the reeling-mechanism's inertia intensifies these oscillations significantly.
The associated pole with this mode shape moves to the left as the deployed length of the tether increases.
The mode shapes associated with the longitudinal waves that travel the tether's length with the speed $ \sqrt{E_tA_t/\mu} $ are also evident in the figure.
Although the oscillations associated with these mode shapes are imperceptible by reeling-speed measurements, they appear in the tether's force and consequently in the generated power. 
Such an effect is reported in the experimental flights carried out by the reference \cite{malz2019reference}.
Another interesting aspect of the system is the stability of the tether's force control without feedback as can be seen from $ U_1 \rightarrow F_t $.
The effect of changing the angle of attack mostly affects the DC gains of the magnitude diagrams. 
\section{Gain-Scheduled $\mathcal{H}_{\infty}$ Control}\label{gss}
In this section, we review the guideline for synthesizing gain-scheduled $\mathcal{H}_{\infty}$ output-feedback controllers for the class of polytopic LPV systems \cite{apkarian2000advanced}.
By this method, we seek a family of linear controllers along with a continuous scheduling algorithm that stabilizes the plant and guarantees the system's performance over the entire parameter space.

\begin{figure}[h]
	\centering
	\includegraphics[width=0.27\linewidth]{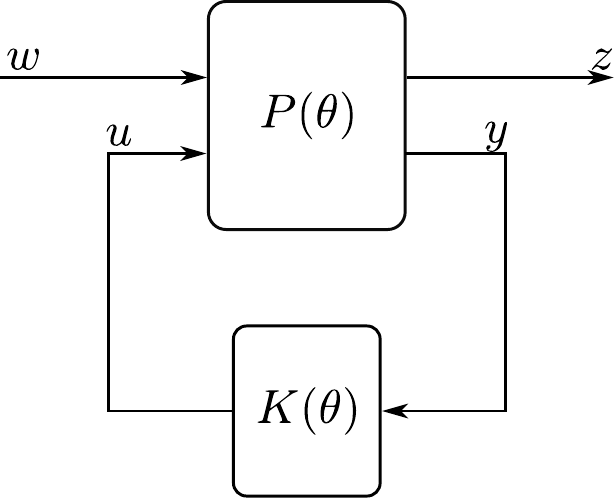}
	\caption{The closed-loop polytopic system}
	\label{fig:pk}
\end{figure}

Consider the LPV system 
\begin{equation}\label{aug}
	P(\theta)\coloneqq\begin{cases}
		\dot{x}_a=A(\theta) x_a+B_{1}(\theta)w+B_{2}u_{},\\
		z=C_{1}(\theta)x_a+D_{11}(\theta)w+D_{12}u,\\
		y=C_{2}x_a+D_{21}w.\\
	\end{cases}
\end{equation}
in which $ x_a $ is the state-vector, $ w $ denotes the vector of disturbances (noise, reference signal, etc.), $ u $ is the control input, $ z $ is the performance output (defined by designer), and $ y $ is the measurable output.
It is assumed that the set of $ r $ time-varying parameters $ \theta(t) = (\theta_1(t),\dots,\theta_r(t)) $ ranges in the known bounded parameter domain $ \Theta\in\mathbb{R}^r $ and can be measured in real-time.
The objective of gain-scheduled $\mathcal{H}_{\infty}$ control is to find a parameter-varying controller in the form of
\begin{equation}\label{controller}
	K(\theta)\coloneqq\begin{cases}
		\dot{x}_k = A_k(\theta)x_k+B_k(\theta)y\\
		u = C_k(\theta)x_k+D_k(\theta)y,
	\end{cases}
\end{equation}
so that the closed-loop system formed by the interconnection of $ P(\theta) $ and $ K(\theta) $ as shown in Fig.~\ref{fig:pk} is internally stable and the maximum induced $ \mathcal{L}_2 $-norm of the operator $ T_{zw}: w\rightarrow z $ is smaller than $ \gamma>0 $, i.e.,
\begin{equation}\label{2norm}
	\int_{0}^{T}z^Tzdt<\gamma^2 \int_{0}^{T}w^Twdt,\quad \forall T.
\end{equation}
For the interconnected system, the maximum gain $ \gamma $ equals the infinity norm of the transfer matrix $ T_{zw} $, i.e.,
\begin{equation}\label{infnorm}
	||T_{zw}(i\omega)||_\infty = \sup_{\substack{\omega,\\\theta\in\Theta}}\bar{\sigma}(T_{zw}(i\omega)),
\end{equation}
where $ \bar{\sigma}(.) $ denotes the largest singular value.

The synthesis of $ K(\theta )$ can be mathematically formulated as convex optimization problem defined by a set Linear Matrix Inequalities (LMIs) in terms of the matrices of the augmented plant $ P_a(\theta) $ as known variables, and the controller system's matrices and two symmetric matrices $ X $ and $ Y $ called Lyapunov variables as unknowns.
Solving the LMI problem is greatly simplified if we assume that the parameter domain $ \Theta $ is a convex polytope with $ m $ vertex and the LPV system is expressed as a convex combination such that 
\begin{equation}\label{affineP}
	\left[\begin{array}{ccc}
		A(\theta)&B_{1}(\theta)&B_{2}\\
		C_1(\theta)&D_{11}(\theta)&D_{12}\\
		C_2&D_{21}&0
	\end{array} \right] = \sum_{i=1}^{m}\rho_i(\theta)
	\left[\begin{array}{ccc}
		A_{i}&B_{1,i}&B_{2}\\
		C_{1,i}&D_{11,i}&D_{12}\\
		C_2&D_{21}&0
	\end{array} \right],\ \forall\theta\in\Theta, 
\end{equation}
where the index $ i $ for $ i=1\ \dots m $ denotes the association with the $ i $-th vertex and $ \rho_i(\theta) $s are scalar basis-functions that satisfy
\begin{equation}\label{rho}
	\sum_{i=1}^{m}\rho_i(\theta)=1,\ \forall\theta\in\Theta.
\end{equation}
Notice the restriction of the dependency on $\theta$ in Eq.~\eqref{affineP}.
In this case, we only need to solve the LMIs for the $ m $ vertex plant systems and can use the same basis-functions to express the parameter-dependent controller over $ \Theta $, i.e., 
\begin{equation}\label{affineK}
	\left[\begin{array}{cc}
		A_k(\theta)&B_{K}(\theta)\\
		C_k(\theta)&D_{K}(\theta)
	\end{array} \right] = \sum_{i=1}^{m}\rho_i(\theta)
	\left[\begin{array}{ccc}
		A_{K,i}&B_{K,i}\\
		C_{K,i}&D_{K,i}
	\end{array} \right].
\end{equation}

To formulate the synthesis problem, we use the scaled version of the basic characterization LMIs.
By this formulation, we seek to minimize the infinity norm of $ ST_{zw}S^{-1} $, where $ S $ is a diagonal scaling matrix.
The scaling reduces the conservatism of the resulting controller by focusing the optimization problem on minimizing the meaningful elements of $ T_{zw} $ \cite{apkarian2000advanced}. 
Moreover, we restrict the regions in which closed-loop poles of the system are placed during the synthesis by appending an additional set of LMIs (\cite{chilali1996h}). 
The complete set of LMIs is provided in \nameref{Appendix}.
\subsection{Transformation to Polytopic Form}
The LPV model of Eq.~\eqref{state_space} is parameterized by the variable set $ \theta = (\bar{f}_{t},\bar{\alpha}, \ell) $ in a nonlinear fashion and cannot be expressed readily in the polytopic form of Eq.\eqref{affineP}.
To remedy this issue, a numerical algorithm known as higher-order singular-value decomposition (HOSVD) can be used.
By using HOSVD, we find a set of basis-functions $ \rho_i(\theta) $ and their associated vertex matrices that approximate the parameter-dependent augmented plant in the polytopic form.

To utilize the HOSVD, we need to supply the analytical form of $ P(\theta) $ along with a sufficiently fine grid over the parameter domain $ \Theta $ to the algorithm provided in the reference \cite{Baranyi_2004}.
The decomposition algorithm returns the set of $ I_1\times I_2,\dots\ \times I_m $ vertex systems $ S $ along with weight functions $ U_{j,i_m} $ that reconstruct the original parameter-dependent matrix as
\begin{equation}\label{hosvd}
	P(\theta(t)) \approx \sum_{i_1=1}^{I_1}\dots\sum_{i_m=1}^{I_m}\prod_{j=1}^{m}U_{j,i_m}(\theta_j(t))S_{i_1,\dots,i_m}
\end{equation}
where $ [I_1,I_2,\dots\ I_m] $ are the number of singular values that correspond to the $ j $-th element of $ \theta(t) $. 
\section{Control strategy}\label{cs}
The proposed control strategy concerns the traction phase of the tethered- wing power system and resembles the pitch-to-feather strategy that is used for variable-speed variable-pitch control of wind turbines (see \cite{bianchi2007control,Poureh_2020} as examples).
To facilitate the description, we introduce the concepts of operational space and available wind power. 
The former is defined as the space formed by the set of variables $ (v_w,v_r,\alpha) $ ranging in their respective span of possible values.
To define the latter, consider the instantaneous power of the tethered-wing system with perfect efficiency as the product of the aerodynamic force and the reeling-speed.
Additionally, assume that the wing's speed in the direction of the tether matches the reeling-speed.
Hence, the instantaneous mechanical power is given by
\begin{equation}\label{power1}
	P =c(\alpha,\ell)(v_{w}\cos\phi-v_r)^2v_r,
\end{equation}
which is maximized for any $ \alpha $ and $ \ell $ at $ v_{r,opt} = \frac{1}{3}v_{w}\cos\phi $ for $ v_r\in[0\ v_w\cos\phi] $.
By substituting $ v_{r,opt} $ in Eq.~\eqref{power1} and assuming that the maximum value for the aerodynamic coefficient $ c(\alpha,\ell) $ is at $ \alpha_{max} $ for any $ \ell $, the available power for wind speed $ v_w $ is
\begin{equation}\label{p_max}
	P_{max} = \dfrac{4}{27}c(\alpha_{max},\ell)v_{w}^3\cos^3\phi.
\end{equation}
The comparison between the available power and generator's rated power is the basis for defining high and low wind speed regions.

\begin{figure}[h]
	\centering
	\includegraphics[width=0.6\linewidth]{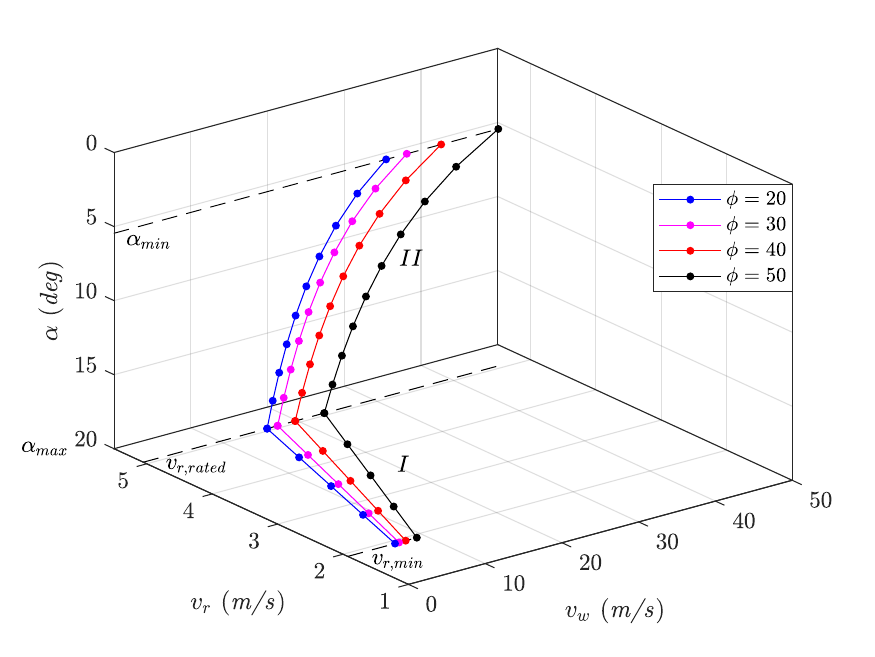}
	\caption{Operating strategy of tethered-wing power system during the reel-out phase for $ \ell=200\ m $}
	\label{fig:powercurve}
\end{figure}

In the low wind speed region, where the available power is less than the rated value, the goal is to convert all of the available wind power to mechanical power.
Therefore, the wing's angle of attack is set to its maximum value and the reeling-speed is controlled to match the theoretical optimal value $ v_{r,opt} $.
Figure~\ref{fig:powercurve} illustrates the low wind speed region as the $ v_wv_r $-plane at $ \alpha_{max} $ denoted by $ I $, in which lines $ v_r=\frac{1}{3}v_{w}\cos\phi $ are displayed for various values of $ \phi $ and $ \ell=200\ m $.
The value $ v_{r,rated} $ corresponds to the rated power and is equal for all $ v_w $ and $ \phi $.
The same is true for $ v_{r,min} $ that is calculated from the minimum generated power. 

The high wind speed region is defined as the subspace in which the available power exceeds the generator's rated value.
In this region, the objective is to maintain the rated value of the generated power and prevent power overload.
This can be achieved by keeping the constant reeling-speed $ v_{r,rated} $ and decreasing the angle of attack as the wind speed increases.
The locus of this region is in the $ v_w\alpha $-plane and is denoted by $ II $ in Fig.~\ref{fig:powercurve}.
Notice that increasing the elevation angle $ \phi $ can also be used to limit the generated power.
However, this cannot be considered as a viable strategy to deal with wind gusts, but rather a method to decrease pitch angle activity.

\begin{figure}[h]
	\centering
	\includegraphics[width=0.5\linewidth]{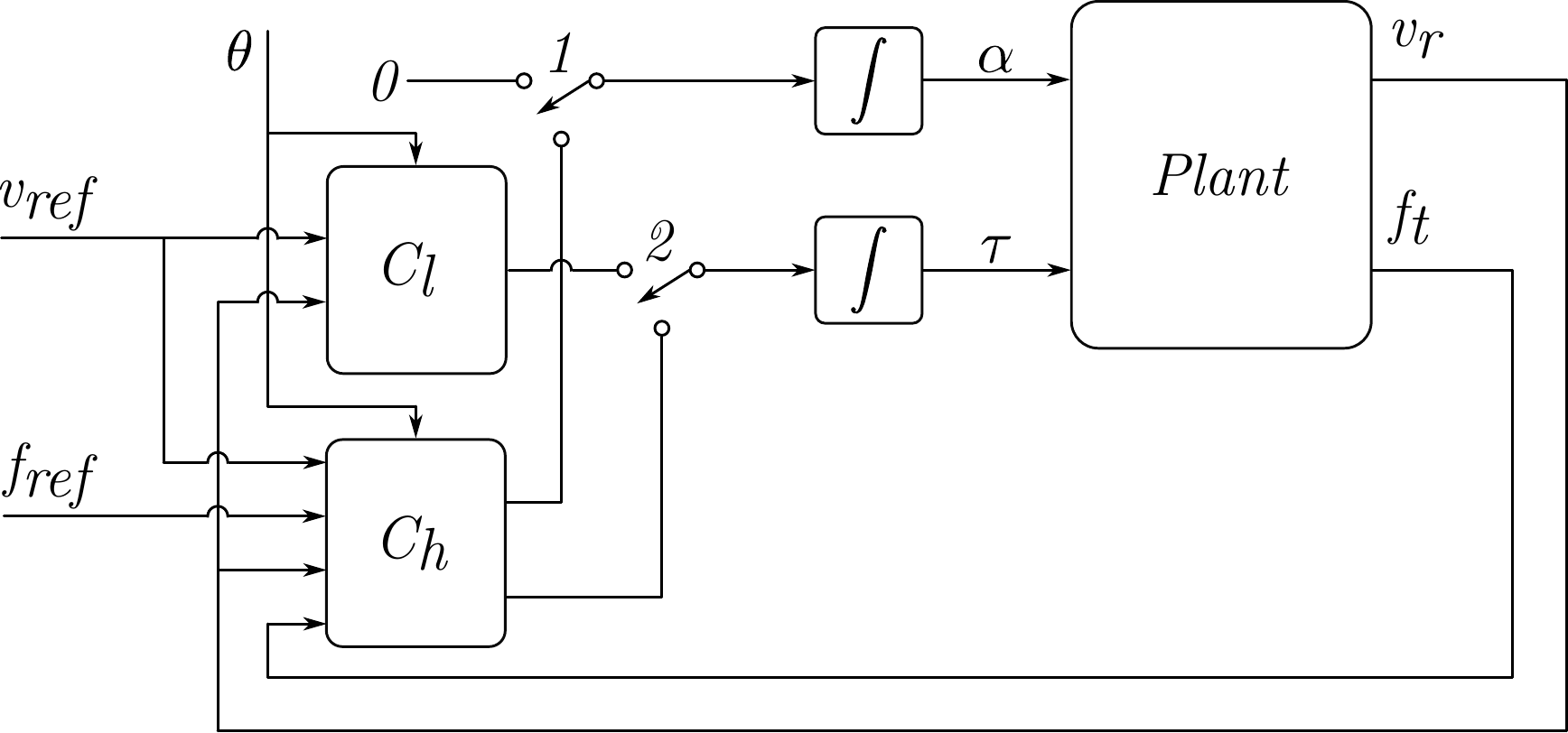}
	\caption{Controller switching}
	\label{fig:strategy}
\end{figure}

We realize the control strategy by using a designated controller for each operational region and switching between them according to the comparison of the rated and available power.
In the low wind speed region, a reeling-speed controller closes the feedback loop from the reeling-speed (controlled variable) to the generator torque (control input).
The reference signal for the controller is chosen so that the reeling-speed tracks the optimal value $ v_{r,opt} $.
In the high wind speed, a multivariable controller is placed in the feedback loop and both the angle of attack and the torque are commanded (control inputs).
The multivariable controller's setpoints are the rated values of torque and reeling speed (controlled variables).
Figure~\ref{fig:strategy} illustrates the feedback loop in which $ C_l $ and $ C_h $ denote the low and high wind speed controllers, respectively.
To smooth the transitions between the controllers, we conveniently place the controllers and the switches before integrators.
This method also obviates the need to supply the steady-state values for the control inputs.

We design both controllers for the entire operational region so that they can be used for the retraction and transitional phases as well.
Admittedly, this will add to the conservatism of the controllers. 
However, we show in the simulations that the controller has satisfactory performance even with the added conservatism.
\section{Robust LPV Gain-Scheduled Control}\label{sec6}
The parameter set $ (\bar{f}_t,\bar{\alpha},\ell) $ determines the operating-point of the tethered-wing system, which suggests its use for scheduling the controllers as discussed in sections \ref{gss} and \ref{cs}.
However, such controllers require the online measurement of the wing's angle of attack $ \alpha $ which may be inaccurate.
Moreover, the aerodynamic model on which the scheduled controller design relies can be subject to significant errors.
To deal with this issue, we propose two different approaches to the gain-scheduling problem.

In the first approach, we consider only the tether's deployed length as the scheduling parameter, i.e., $ \theta = (\ell) $,
and deal with the variations of $ \bar{f}_t $ and $ \bar{\alpha} $ as parametric uncertainties. 
In the second approach, it is assumed that the parameter set is available to schedule the controller, i.e., $ \theta = (\bar{f}_t,\bar{\alpha},\ell) $.
However, we still include the parametric uncertainty of the scheduling variables $ \bar{f}_t $ and $ \bar{\alpha} $ to deal with the aerodynamic model mismatch and the measurement errors.

To express the parameter uncertainties in the LPV plant, we substitute the instances of $ \sqrt{c(\bar{\alpha},\ell)\bar{f}_t} $ and $ \frac{c_\alpha(\bar{\alpha},\ell)}{c(\bar{\alpha},\ell)}\bar{f}_t $ in Eq.~\eqref{hatf_w2} with $ p_1 +\delta_1 p_{d,1} $ and $ p_2 +\delta_2 p_{d,2} $, respectively, where $ p_i $ denotes the nominal values, $ p_{d,i} $ is the magnitude of variations with respect to the nominal values, and the real scalar $ |\delta_i|<1 $ is the uncertain parameter.
Using the linear fractional transformation, the outputs associated with the parameter uncertainties are found as $	z_{\delta,1} = \dot{\hat{q}}_{n+1} $ and $ \ z_{\delta,2} = \hat{\alpha} $.
Constructing Eq.~\eqref{state_space} for $ n=1 $, we express the reduced-order dynamics with parametric uncertainty as
\begin{equation}\label{lpv1}
	P(\theta):=\begin{cases}
		\dot{\hat{x}} = A(\theta)\hat{x} + B_1(\theta) \hat{v}_w+ B_d(\theta) w_\delta+ B_2(\theta) \hat{u},\\
		z_\delta =C_\delta \hat{x}+B_\delta \hat{u}\\
		\hat{y} = C_m\hat{x},
	\end{cases} 
\end{equation}
in which
\begin{equation}\label{b1}
	\begin{gathered}
		A(\theta)=\dfrac{1}{m_3}\left[\begin{array}{ccc}
			0 & -\frac{m_3}{\ell}                & \frac{m_3}{\ell}                                        \\
			(m_1 + \frac{\mu\ell}{6})EA & m_1(b_d+b_t) + \frac{\mu\ell}{6}b_t   & -m_1b_t-\frac{\mu\ell}{6}(b_t + 2p_1) \\
			-(m_2 + \frac{\mu\ell}{6})EA & - m_2b_t - \frac{\mu\ell}{6}(b_d+b_t) & \frac{\mu\ell}{6}b_t+ m_2(b_t + 2p_1)
		\end{array} \right],\\
		B_1(\theta) = \dfrac{1}{m_3}\left[\begin{array}{c}
			0\\\frac{\mu\ell}{3}p_1\\-2m_2p_1
		\end{array} \right],\
		B_2(\theta)= \dfrac{1}{m_3}
		\left[\begin{array}{cc}
			0                                & 0                                                                    \\
			m_1\frac{\eta}{r}                & \frac{\mu\ell}{6}p_2 \\
			-\frac{\mu\ell}{6}\frac{\eta}{r} & m_2p_2
		\end{array} \right],\\
		B_d(\theta)=\dfrac{1}{m_3}\left[\begin{array}{cc}
			0&0\\
			\frac{\mu\ell}{3}p_1&\frac{\mu\ell}{6}p_2\\
			-2m_2p_1&m_2p_2
		\end{array} \right], B_{\delta} =  \left[\begin{array}{cc}
			0&0\\0&1
		\end{array} \right]	\\
		C_{\delta} = \left[ \begin{array}{ccc}
			0&0&1\\0&0&0
		\end{array}\right], C_m = \left[\begin{array}{ccc}
			0&1&0\\EA&0&0
		\end{array} \right],
	\end{gathered}
\end{equation}
and where $ m_1 = m_k + \mu\ell/3 $, $ m_2 = m_r + \mu\ell/3 $,  $ m_3 = m_km_r + \mu\ell(m_r+ m_k)/3 + \ell^2\mu^2/12 $, the state-vector is $ \hat{x}=\left[\hat{\epsilon}_1\ \dot{\hat{q}}_1\ \dot{\hat{q}}_2 \right]^T $, and $ w_{\delta}=[w_{\delta,1}\ w_{\delta,2}]^T $. 
By the application of the small-gain theorem, the stability of the uncertain system is guaranteed if the maximum gain across the channels $ w_{\delta,i}\rightarrow z_{\delta,i} $ is less than $ 1 $.

\begin{figure}[h]
	\centering
	\begin{subfigure}[b]{0.3\textwidth}
		\centering
		\includegraphics[width=\textwidth]{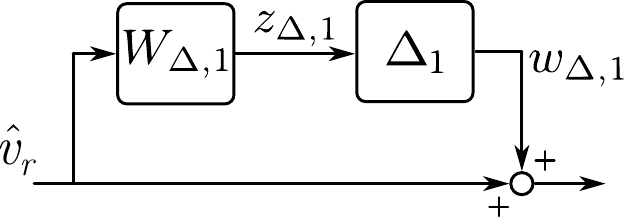}
		\caption{Reeling-speed uncertainty}
		\label{fig:output-multiplicative-uncertainty-1}
	\end{subfigure}\qquad\qquad
	\begin{subfigure}[b]{0.3\textwidth}
		\centering
		\includegraphics[width=\textwidth]{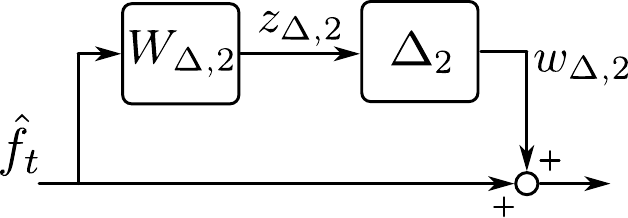}
		\caption{Tether force uncertainty}
		\label{fig:output-multiplicative-uncertainty-2}
	\end{subfigure}
	\caption{Output multiplicative uncertainties}
	\label{fig:uncertainty}
\end{figure}

Another source of uncertainty in the LPV model is due to the fact that the reduced-order dynamic model Eq.~\eqref{lpv1} leaves out the tether's longitudinal vibrational modes and represents only the coupled oscillations of the tether and the drum.
To deal with this, we consider output multiplicative uncertainties as shown in Fig.~\ref{fig:uncertainty} such that
\begin{equation}\label{W}
	|W_{\Delta,i}(j\omega)|\geq \sup_{\substack{\omega,\\\theta\in\Theta}}\bar{\sigma}|G_{r,i}(j\omega,\theta)-G_i(j\omega,\theta)|,
\end{equation}
in which $ G_{i}(j\omega,\theta) $ is the frequency response of the $i$-th row of the transfer matrix Eq.~\eqref{tf_ol} and $ G_{r,i}(j\omega,\theta) $ represents that of the reduced-order dynamic model.
The uncertain operators $ \Delta_i $ are defined as $ \left\lbrace \Delta_i \in \mathscr{C}_{LTI}: ||\Delta_i||_\infty<1\right\rbrace $, in which $ \mathscr{C}_{LTI} $ denotes the set of all stable LTI plants. In such a case, the small-gain theorem ensures the stability of the uncertain system if the maximum gain across $ w_{\Delta,i}\rightarrow z_{\Delta,i} $ channels is less than $ 1 $.
\subsection{Reeling-speed control}
\begin{figure}[h]
	\centering
	\includegraphics[width=.6\linewidth]{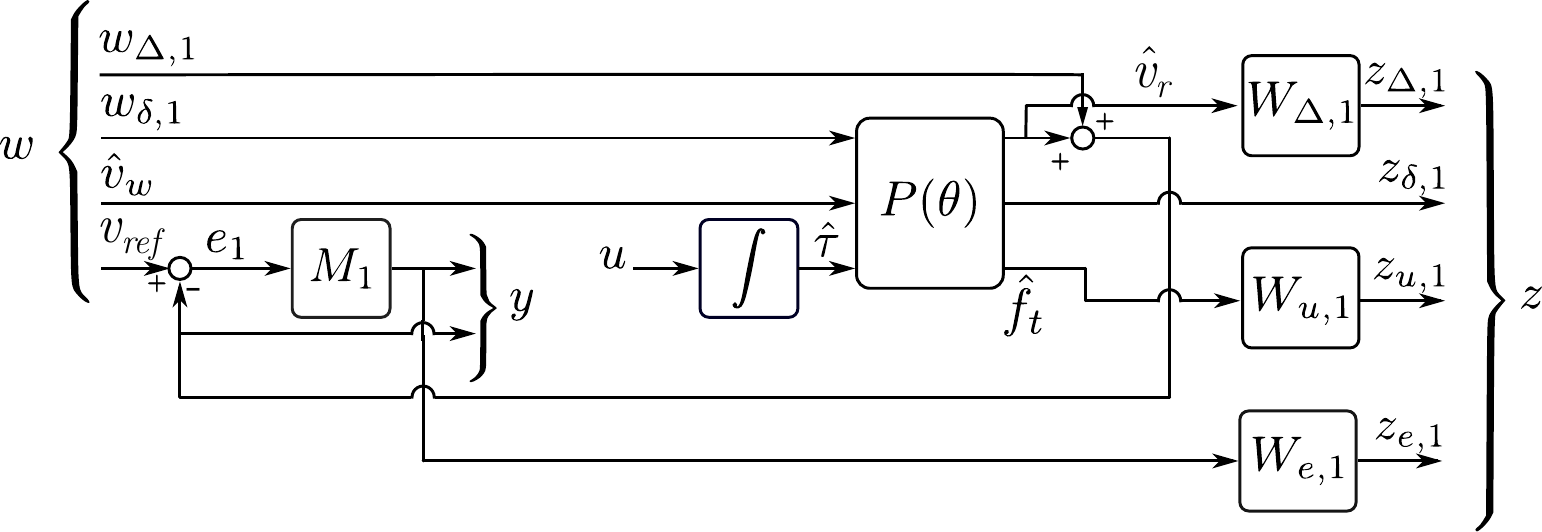}
	\caption{Augmented plant for speed control (1-DoF)}
	\label{fig:augmented-plant-1dof}
\end{figure}

The reeling-speed controller, henceforth referred to as 1-DoF controller, has the control signal $ \dot{\tau} $ and controlled variable $ v_r $. 
The augmented plant for the reeling-speed controller plant is formed with the following objectives:
\begin{enumerate}
	\item Smooth reeling-speed control with sufficient bandwidth;
	\item Minimize tether's force over-shoots;
	\item Being robust with respect to the dynamic and parametric uncertainties.
\end{enumerate}
Figure~\ref{fig:augmented-plant-1dof} illustrates the augmented plant's block diagram for the speed control scheme in which $ v_{ref} $ is the reference reeling-speed, and $ M_1 $ is a low-pass filter added to make the speed control loop stabilizable.
The weighting functions are presented in section \ref{csy}.
\subsection{Reeling-speed and tether force control}
\begin{figure}[h]
	\centering
	\includegraphics[width=.81\linewidth]{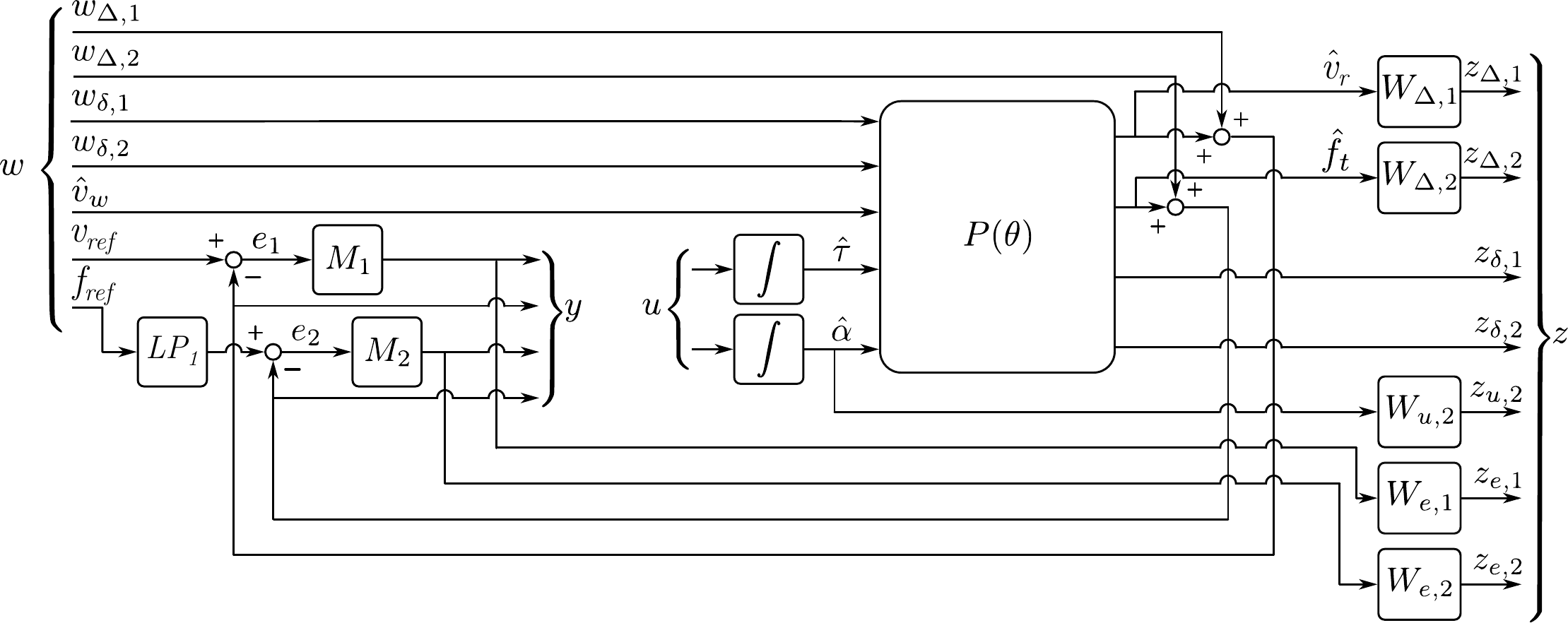}
	\caption{Augmented plant for the multivariable control (2-DoF)}
	\label{fig:augmented-plant-2dof}
\end{figure}

The multivariable controller, henceforth referred to as 2-DoF controller, has the control inputs $ \dot{\tau} $ and $ \dot{\alpha} $ and controlled variables $ v_r $ and $ f_t $ and is synthesized without any allocation of control inputs to the controlled outputs.
We take advantage of this degree of freedom to reduce the pitch angle activity by penalizing the fast variations of the angle of attack.
In this case, the controller objectives are:
\begin{enumerate}
	\item Speed and force controller with sufficient bandwidth;
	\item Minimize the bandwidth of the demanded angle of attack;
	\item Being robust with respect to the dynamic and parametric uncertainties.
\end{enumerate}
The weighting functions used to achieve the objectives are presented in section \ref{csy}.
Figure~\ref{fig:augmented-plant-2dof} shows the augmented plant for multivariable control, in which $ f_{ref} $ is the tether's force reference signal.
Similar to the reeling-speed control, $ M_2 $ is added to make the force control loop stabilizable.
\subsection{Controller Synthesis}\label{csy}
The augmented plants of Figs.~\ref{fig:augmented-plant-1dof} and \ref{fig:augmented-plant-2dof} were formed with the reduced-order dynamic model $ (n=1) $ and the parameter values shown in Table \ref{table:param}. 
The weighting functions were chosen as
\begin{align}
	M_1W_{e,1}&=\dfrac{0.8s+1}{s}, & M_2W_{e,2}&=\dfrac{0.1s+1}{s}, \\
	W_{u,1}&=\dfrac{4s}{s+20}, & W_{u,2}&=\dfrac{s}{s+1},\\
	W_{\Delta,1}&=\dfrac{(3.16s+0.63)^2}{(s+63.24)^2}, & W_{\Delta,2}&=\dfrac{(3.16s+0.63)^2}{(s+63.24)^2},\\LP_1 &= \dfrac{1}{s+1},&&
\end{align}
where $ W_{e,1}=.75 $ and $ W_{e,2}=1 $.
The weighting functions $ M_1W_{e,1\& 2} $ are chosen with a pole at the origin to ensure an integral action in the controllers. 
Their bandwidths is chosen based on how fast we wanted the system to react to a change of the reference values.
The weighting functions $ W_{u,1\& 2} $ penalize the high frequencies of the tether's force and the angle of attack. 
The choice of $ W_{\Delta,1\& 2} $ were made to satisfy the condition Eq.~\eqref{W}.
With this condition, it is guaranteed that the resonance frequencies of the system are attenuated by passing through the controller.
Finally, the filter $ LP_1 $ is added to reduce the conservatism of the synthesized controller with respect to the reference signal $ f_{ref} $ by only allowing its lower frequencies to pass.
The exact values of the weighting functions, except for $ W_{\Delta,1\& 2} $, were chosen by trial and error until satisfactory results were achieved.

The HOSVD algorithm was used to express the augmented plants in the polytopic form by $ 12 $ and $ 27 $ vertices for the 1-DoF and 2-DoF cases with three scheduling parameters, respectively.
In the case of single scheduling parameter ($\theta = (\ell)$), the plant was reconstructed by $ 3 $ vertices.
We chose the parameters' domain as $ f_t\in[1.5\ 20]\ KN $, $ \alpha\in[4\ 20]\ \deg $, and $ \ell\in[200\ 800]\ m $ and formed a $ 50 $ point grid for each dimension.
The operating point for the single scheduling parameter controllers were chosen as $ \bar{f}_t= 10\ kN $ and $ \bar{\alpha}= 10\ \deg $.

\begin{figure}[h]
	\centering
	\begin{subfigure}[b]{0.35\textwidth}
		\centering
		\includegraphics[width=\textwidth]{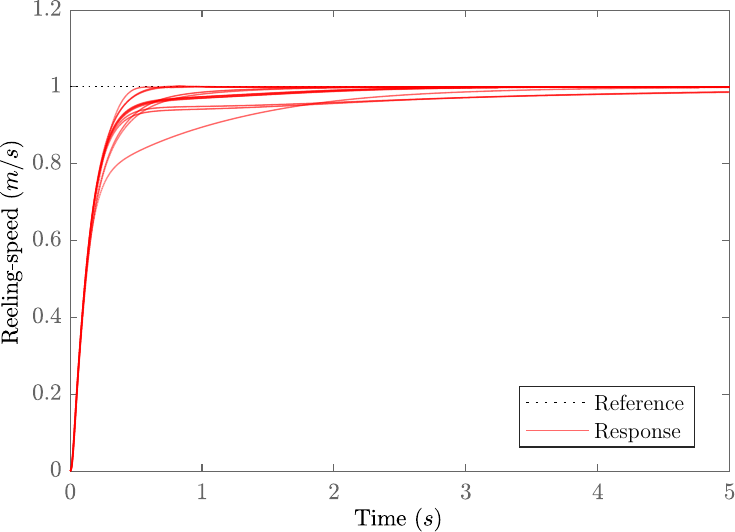}
		\caption{Scheduled with $ \theta =(\bar{f}_t,\bar{\alpha},\ell) $}
		\label{fig:00-speed}
	\end{subfigure}\qquad\qquad
	\begin{subfigure}[b]{0.35\textwidth}
		\centering
		\includegraphics[width=\textwidth]{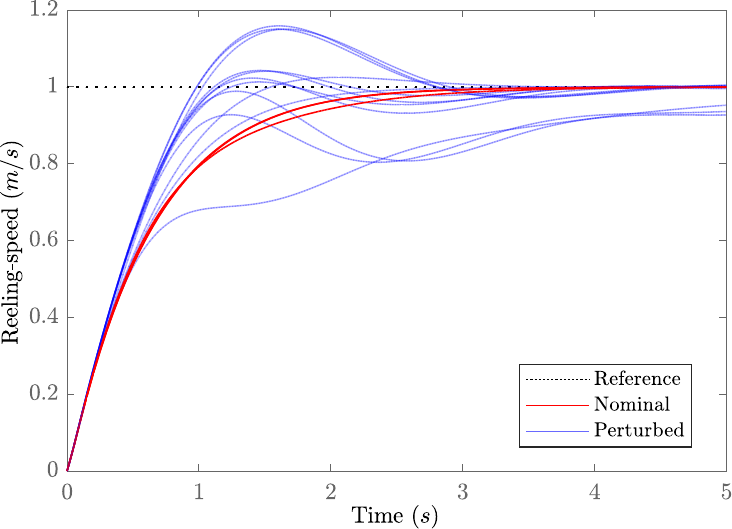}
		\caption{Scheduled with $ \theta =(\ell) $}
		\label{fig:00-speed-robust}
	\end{subfigure}
	\caption{Step response of 1-DoF controllers}
	\label{fig:1dof}
\end{figure}

\begin{figure}[h]
	\centering
	\begin{subfigure}[b]{0.35\textwidth}
		\centering
		\includegraphics[width=\textwidth]{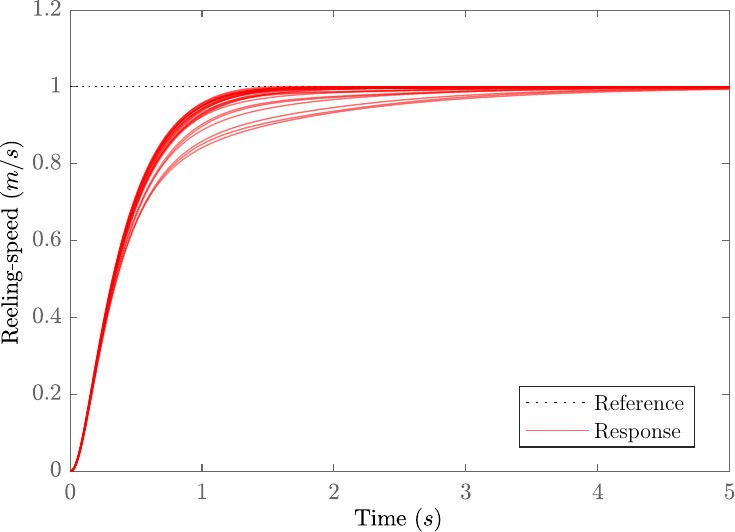}
		\caption{Reeling-speed response}
		\label{fig:00-speed-2dof}
	\end{subfigure}\qquad\qquad
	\begin{subfigure}[b]{0.35\textwidth}
		\centering
		\includegraphics[width=\textwidth]{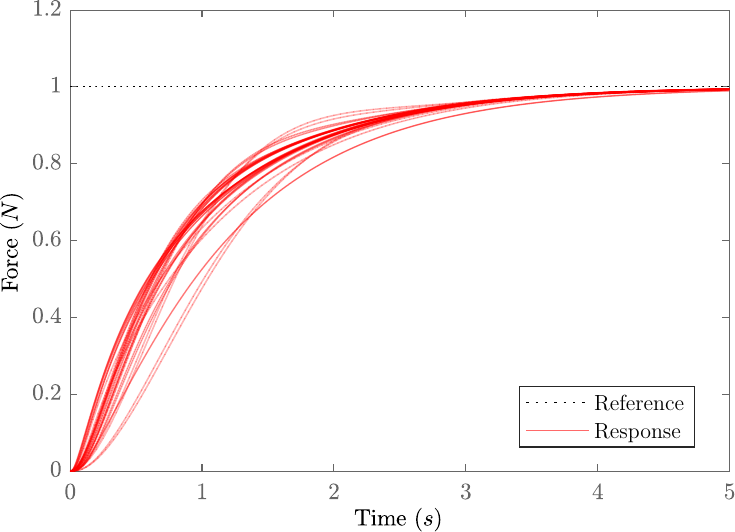}
		\caption{Force response}
		\label{fig:00-force-2dof}
	\end{subfigure}
	\caption{Step response of 2-DoF controller scheduled with $ \theta =(\bar{f}_t,\bar{\alpha},\ell) $}
	\label{fig:2dof}
\end{figure}

\begin{figure}[h]
	\centering
	\begin{subfigure}[b]{0.35\textwidth}
		\centering
		\includegraphics[width=\textwidth]{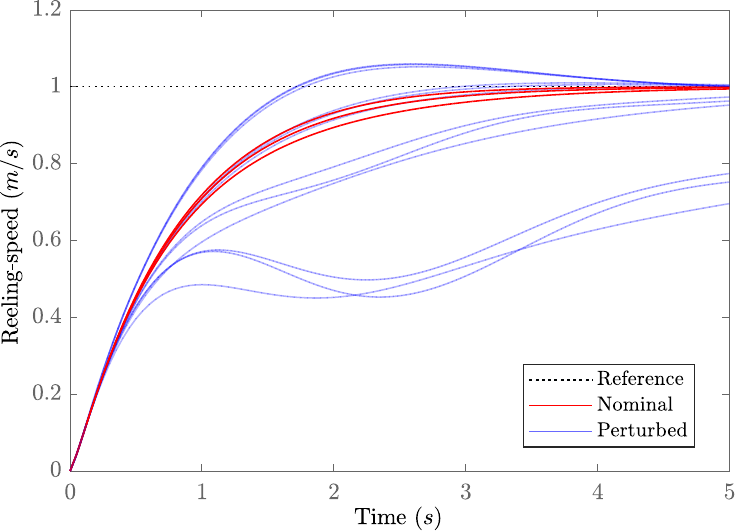}
		\caption{Reeling-speed response}
		\label{fig:00-speed-2dof-robust}
	\end{subfigure}\qquad\qquad
	\begin{subfigure}[b]{0.35\textwidth}
		\centering
		\includegraphics[width=\textwidth]{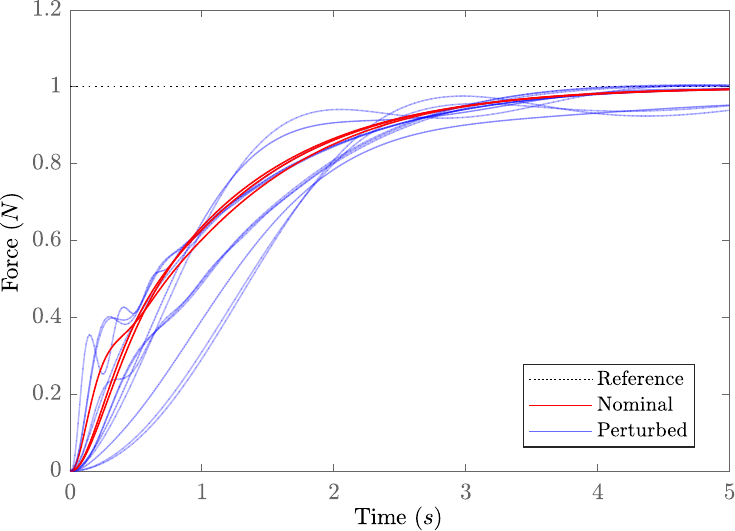}
		\caption{Force response}
		\label{fig:00-force-2dof-robust}
	\end{subfigure}
	\caption{Step response of 2-DoF controller scheduled with $ \theta =(\ell) $}
	\label{fig:2dof-robust}
\end{figure}

To synthesize the controllers, we carried out the basic characterization procedure with different choices for constant Lyapunov variables $ X $ and $ Y $ and chose the controller with the least $ \gamma $.
The poles of the closed-loop system for the 1-DoF and 2-DoF cases were constrained to the vertical stripes of the complex plane $ |Re(z)|<200 $ and $ |Re(z)|<400 $, respectively.
The step responses of the controllers are shown in Figs.~\ref{fig:1dof}-\ref{fig:2dof-robust}, in which the superiority of the controllers with three scheduling variables is evident.
\section{Simulations}\label{sec7}
\subsection{Full-order Model}
\begin{figure}[h!]
	\centering
	\includegraphics[width=0.5\linewidth]{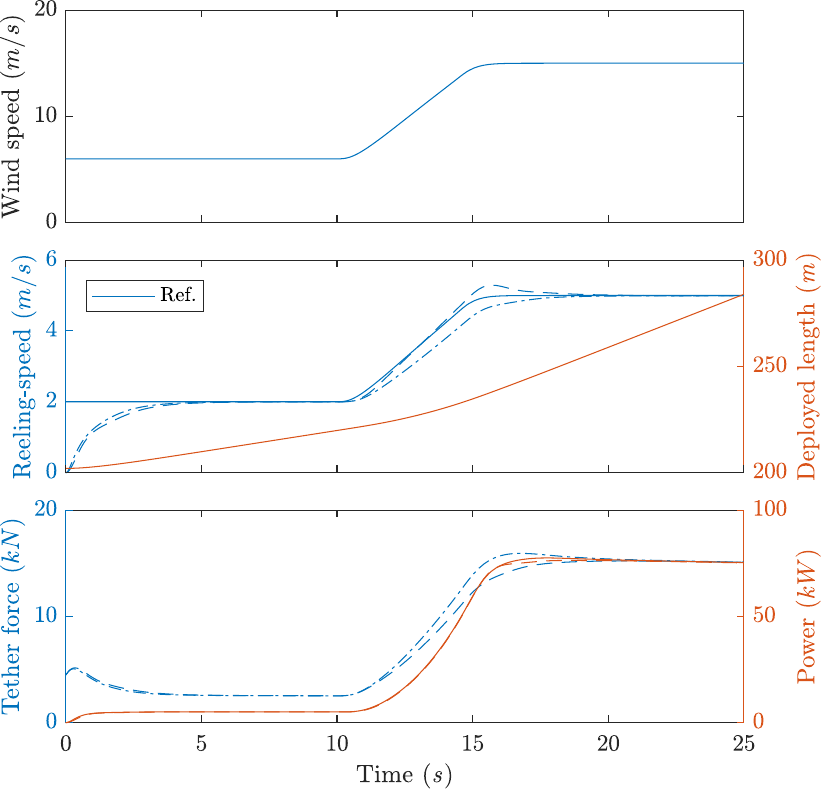}
	\caption{The performance of the 1-DoF control $ (\alpha=18\ \deg) $. Dashed lines: controller scheduled with $  \theta =(\bar{f}_t,\bar{\alpha},\ell) $, dash-dotted line controller scheduled with $ \theta = (\ell) $}
	\label{fig:performance1dof}
\end{figure}

\begin{figure}[h!]
	\centering
	\includegraphics[width=0.5\linewidth]{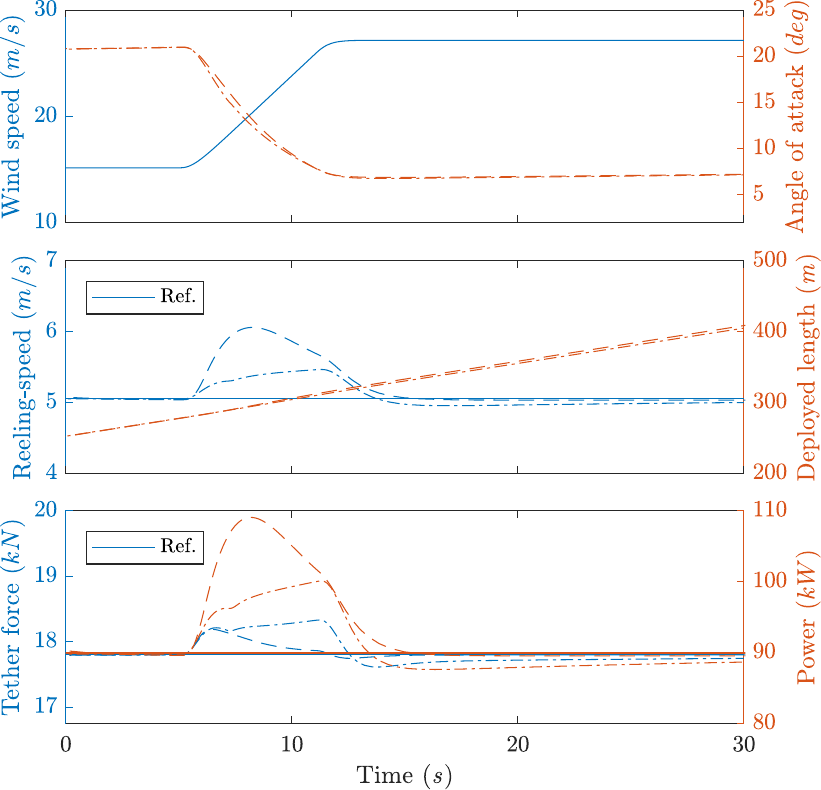}
	\caption{The performance of the 2-DoF control. Dashed lines: controller scheduled with $  \theta =(\bar{f}_t,\bar{\alpha},\ell) $, dash-dotted line controller scheduled with $ \theta = (\ell) $}
	\label{fig:performance2dof}
\end{figure}

To study the performance of the controllers in tracking a reference signal, we connected them to the full-order model with five nodes, in which the tether's length was varied by integrating the reeling-speed.
In both cases, the controllers were scheduled with $ 30\% $ error in the scheduling variables. 
Figure~\ref{fig:performance1dof} illustrates the controlled system behavior for the 1-DoF controllers, in which the directional wind speed varies in the span of the first region of operation and the reeling-speed's setpoint is chosen as the optimal reeling-speed $ v_{r,opt} $ with a priori knowledge of wind speed.
This artificial change of wind speed is designed to illustrate the behavior of the closed-loop system and does not represent a realistic wind scenario.
The angle of attack $ \alpha $ is set to $ 18\ deg $.
The results for the simulation of 2-DoF control are shown in Fig.~\ref{fig:performance2dof}, in which the setpoints are $ v_{ref}=5.05\ m/s $ and $ F_{r}=17.8\ kN$.
In both 1-DoF and 2-DoF controllers, the controller scheduled with $  \theta =(\bar{f}_t,\bar{\alpha},\ell) $ shows better performance compared to the controller with only the deployed tether's length as the scheduling variable. 
\subsection{Simulation in Realistic Wind}
\begin{figure}[h!]
	\centering
	\includegraphics[width=0.45\linewidth]{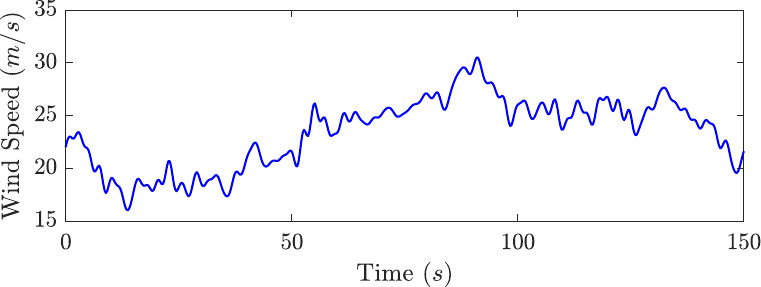}
	\caption{The turbulent wind speed at the wing's height}
	\label{fig:00-cyclewindspeed}
\end{figure}

The control strategy was implemented in a simplified form of the 3-dimensional simulator provided in \cite{Kakavand_2021,Kakavand_2020}.
The simulator represents the tether with $ 4 $ variable-length elastic elements and the wing as a point-mass. 
The generator's dynamic model is not present in the simulator.
The simulation was carried out for a complete cycle in the presence of a turbulent wind field environment that is created according to the Kaimal spectral density \cite{Slootweg_2003} and the logarithmic law.
The flight of the wing along the lemniscate pattern during the traction phase is generated by controlling the wing's velocity angle according to the scheme proposed in \cite{fagiano2013automatic}.
To fly the wing during retraction we have used the regularized velocity angle algorithm provided by \cite{zgraggen2015automatic}.
The wind speed at the wing's height is shown in Fig.~\ref{fig:00-cyclewindspeed}.

\begin{figure}[h!]
	\centering
	\includegraphics[width=0.35\linewidth]{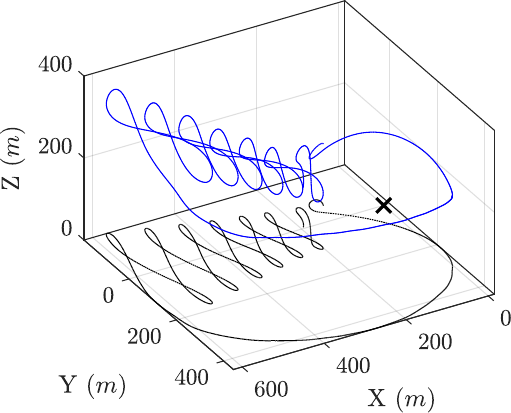}
	\caption{The wing's flight path for a complete cycle. The 'x' mark denotes the origin}
	\label{fig:00-cycle}
\end{figure}

\begin{figure}[h!]
	\centering
	\includegraphics[width=0.45\linewidth]{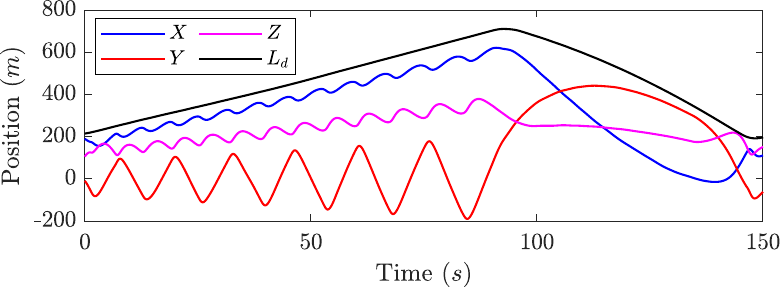}
	\caption{The position of the wing and the deployed tether's length}
	\label{fig:00-cyclexyzl}
\end{figure}

Figure \ref{fig:00-cycle} illustrates the wing's flight path in the $ XYZ $ coordinates, in which the height from the ground and the wind direction are indicated by $ Z $ and $ X $ axes, respectively.
The position of the wing and the deployed length of the tether are shown in Fig.~\ref{fig:00-cyclexyzl}.
Note that the system is in the traction phase for the first $ 90 $ seconds, and after finishing retraction starts a new cycle at $ 145 $ seconds.

\begin{figure}[h!]
	\centering
	\includegraphics[width=0.45\linewidth]{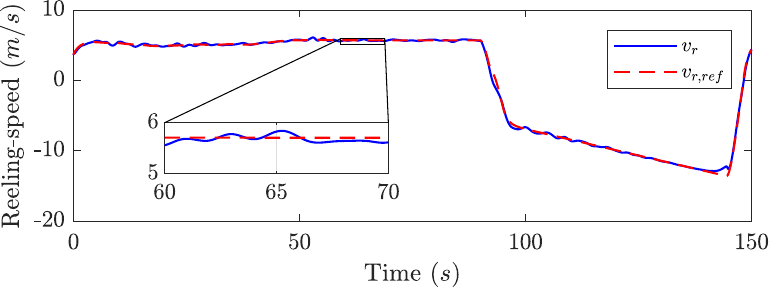}
	\caption{The reeling-speed and its reference signal}
	\label{fig:00-cycle-speed}
\end{figure}

\begin{figure}[h!]
	\centering
	\includegraphics[width=0.45\linewidth]{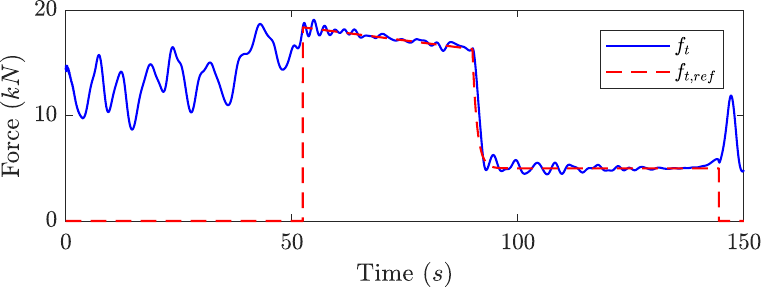}
	\caption{The tether's force and its reference signal. A zero reference signals indicates the use of 1-DoF control}
	\label{fig:00-cycleforce}
\end{figure}

The tether's reeling-speed and its reference signal are shown in Fig.~\ref{fig:00-cycle-speed}.
The deviations from the reference value are due to variations of the wind and the effect of the wing's flight pattern.
The reference signal for the reeling-speed controller was chosen according to the scheme suggested by the reference \cite{erhard2015flight} and without measuring the absolute wind speed.
Figure~\ref{fig:00-cycleforce} illustrates the tether's force and its reference signal, in which a zero reference signal indicates that the speed controller is active.
The switching from low wind speed to high wind speed controllers occurs around $ 53 $ seconds as the generated power reaches $ 110\ kW $.
To make the switching smooth, we use the current values of speed and force as the controller's setpoint at the time of switching.
From that point on, we reduce the force setpoint to compensate for the reduction of the aerodynamic efficiency as the tether's length increases. 
Without this scheme, the commanded angle of attack keeps increasing until it goes out of the acceptable bound.
The use of the multivariable controller has reduced the variations in the tether's force.

\begin{figure}[h!]
	\centering
	\includegraphics[width=0.45\linewidth]{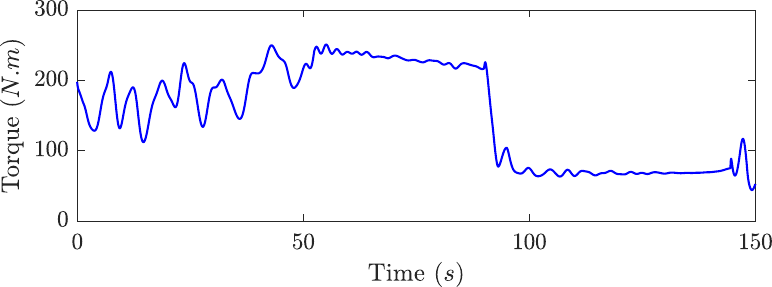}
	\caption{The generator's torque}
	\label{fig:00-cycletorque}
\end{figure}

\begin{figure}[h!]
	\centering
	\includegraphics[width=0.45\linewidth]{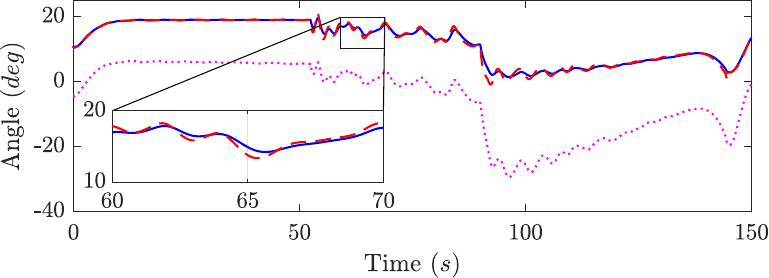}
	\caption{Angle of attack (continuous line), reference signal (dashed line), and the pitch angle (dotted line)}
	\label{fig:00-cycle-angle}
\end{figure}

Figures~\ref{fig:00-cycletorque} and \ref{fig:00-cycle-angle} illustrate the control inputs of the system during the simulation. 
An important simplifying assumption for the controller design was that the angle of attack can be commanded to the wing, while in the real system the angle of attack is determined as a function of the wing's pitch angle and other factors.
It is a reasonable simplification given that the difference between the actual angle of attack and the command can be covered by the parameter uncertainties $ p_{d,1\&2} $. 
In the simulation, this simplification is dealt with by utilizing a PI controller that adjusts the pitch angle based on the difference in the measured value of the angle of attack and the value issued by the controller.
The setpoint for the PI controller, while the speed controller is active, is $ 20\ deg $, which is the maximum value for the angle of attack.

\begin{figure}[h!]
	\centering
	\includegraphics[width=0.45\linewidth]{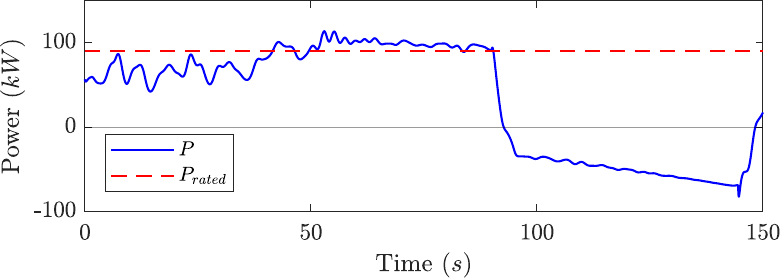}
	\caption{The mechanical power}
	\label{fig:00-cyclepower}
\end{figure}

Finally, Fig.~\ref{fig:00-cyclepower} shows the generated mechanical power computed as the product of the generator's torque and the reeling-speed, in which the positive sign indicates generation. It is evident that the fluctuation in the power has decreased significantly as the multivariable controller is activated.
\section{Conclusions}\label{sec8}
The power regulation of tethered-wing systems is a multi-objective control problem with a critical role in the safety and efficiency of the system.
In that regard, the contribution of this paper is the application of gain-scheduling $ \mathcal{H}_\infty $ and multivariable control for realizing a variable-speed variable-pitch strategy that covers the entire range of operational wind speeds.
This method allows for mitigation of high-frequency mechanical loads while tracking the optimal reeling speed, and minimizing the pitch angle activity while limiting the generated power.
A model for the longitudinal dynamics of the system is also developed that is used to design the controllers and to study the open-loop behavior of the system.
This model characterizes the first mode shape as an out-of-phase motion of the wing and the reeling-drum whose damping ratio decreases with increasing the tether's force and frequency decreases as the deployed length increases.
The effectiveness of the control scheme in executing the full traction-retraction cycle was shown by 3-dimensional simulations in the presence of realistic wind.
It was also demonstrated that scheduling the controllers according to the measurements of aerodynamic coefficient yields considerably better performance compared to making the controllers robust with respect to the variation of the aerodynamic parameters. 
\section*{Appendix}\label{Appendix}
\subsection{Parameters and Constants}
The parameters used for simulation of the tethered-wing system is shown in Table \ref{table:param}.
The physical parameters of the tether are chosen to reflect those of a UHMWPE cables which are usually used for kite-generators.
\begin{table}[h]
	\centering
	\caption{Simulation parameters}
	\label{table:param}
	\begin{tabular}{lrlr}
		Parameter   &       Value & Parameter &            Value \\ \hline
		$ d_t $     &   $ 4\ mm $ & $ \mu $   & $ 0.0123\ kg/m $ \\
		$ E_t $     & $ 90\ GPa $ & $ L_t $   &       $ 800\ m $ \\
		$ \eta $    &      $ 15 $ & $ I_r $   &   $ 30\ kg.m^2 $ \\
		$ r $     &       $ 0.2\ m $& $ b_d $   &     $ 2\ N.s/m $ \\
		$ C_{D,t} $ &    $ 0.96 $ &$ S_w $     & $ 12\ m^2 $   \\
		$ m_w $     &  $ 29\ kg $ &           &                  \\ \hline
	\end{tabular}
\end{table}

The matrices of the LPV system Eq.~\eqref{state_space} are as follows.
\begin{equation}\label{M}
	M = \left[\begin{array}{ccccc}
		m_r+\frac{1}{3}\mu\ell &\frac{1}{6}\mu\ell & 0 & \dots& 0\\
		\frac{1}{6}\mu\ell &\frac{2}{3}\mu\ell &\frac{1}{6}\mu\ell &0& \vdots\\
		0&\ddots& \ddots& \ddots& 0\\
		\vdots&0& \frac{1}{6}\mu\ell &\frac{2}{3}\mu\ell &\frac{1}{6}\mu\ell\\
		0& \dots&0&\frac{1}{6}\mu\ell&m_k+\frac{1}{3}\mu\ell		
	\end{array} \right]_{(n+1)\times (n+1)},
	{C}=\left[\begin{array}{ccccc}
		b_t & -b_t &0& \dots & 0\\
		-b_t&2b_t& -b_t& 0& \vdots\\
		0&\ddots&\ddots&\ddots&0\\
		\vdots&0&-b_t&2b_t&-b_t\\
		0 & \dots& 0& -b_t&b_t
	\end{array} \right]_{(n+1)\times (n+1)} ,
\end{equation}
\begin{equation}\label{K}
	{K}_\epsilon=\left[\begin{array}{cccc}
		-EA & 0  & 0\\
		EA&\ddots&0\\
		0&\ddots&-EA\\
		0 & 0& EA
	\end{array} \right]_{(n+1)\times n},  L = \left[\begin{array}{cccc}
		-\frac{1}{\ell}&\frac{1}{\ell}&0&0\\
		0&\ddots&\ddots&0\\
		0&0&-\frac{1}{\ell}&\frac{1}{\ell}
	\end{array} \right]_{n\times(n+1)}.
\end{equation}
\begin{equation}\label{linmat}
	C_f = \left[\begin{array}{ccc}
		b_d & 0_{1\times (n-1)} &0\\
		0_{n\times 1} & 0 & 0_{n\times 1}\\
		0 & 0 & 2\sqrt{c(\bar{\alpha},\ell)\bar{f}_t}
	\end{array} \right],
\end{equation}
\begin{equation}\label{key}
	B_w=\left[\begin{array}{c}
		0_{n\times 1}\\2\sqrt{c(\bar{\alpha},\ell)\bar{f}_t}
	\end{array} \right],\ 
	B_\tau = \left[\begin{array}{c}
		-\dfrac{\eta}{r}  \\
		0_{n\times 1} \end{array} \right],\\
	B_\alpha = \left[\begin{array}{c}
		0_{n\times 1}      \\\dfrac{c_\alpha(\bar{\alpha},\ell)}{c(\bar{\alpha},\ell)}\bar{f}_t
	\end{array} \right] 
\end{equation}
\begin{equation}\label{ba}
	C_m = \left[\begin{array}{ccc}
		0_{1\times n}&1&0_{1\times n}\\
		EA&0&0
	\end{array} \right] 
\end{equation}
\subsection{LMIs}
The scaled version of basic characterization LMIs are 
\begin{equation}\label{basic_char}
	\left[ \begin{array}{cccc}
		XA+\tilde{B}C_2 +\star & \star& \star& \star\\
		\tilde{A}^T+A+B_2\tilde{D}C_2 & AY+B_2\tilde{C} +\star& \star& \star\\
		S(XB_1+\tilde{B}D_{21})^T & S(B_1+B_2\tilde{D}D_{21})^T & -\gamma S & \star\\
		C_1+D_{12}\tilde{D}C_2 & C_1Y+D_{12}\tilde{C}&(D_{11}+D_{12}\tilde{D}D_{21})S & -\gamma S 
	\end{array}\right] <0,
\end{equation}
\begin{equation}\label{lyap}
	\left[\begin{array}{cc}
		X & 0\\ 0& Y
	\end{array} \right]>0,
\end{equation}
in which $ \star $ denotes symmetry. For example, $ A+\star= A+A^T $. The controller's matrices are given by
\begin{equation}\label{Ak}
	\begin{split}
		A_k =& N_f^{-1}(\tilde{A}-X(A-B_2\tilde{D}C_2)Y-\tilde{B}C_2Y-XB_2\tilde{C})M_f^{-T},\\B_k =& N_f^{-1}(\tilde{B}-XB_2\tilde{D}),\\
		C_k =& (\tilde{C}-\tilde{D}C_2Y)M_f^{-T},\\D_k =&\tilde{D},
	\end{split}
\end{equation}
where $ M $ and $ N $ are the solution to the factorization problem
\begin{equation}\label{MN}
	I-XY=NM^T.
\end{equation}
To enforce the regional pole constrains, we appended the following LMIs to the basic characterization LMIs
\begin{equation}\label{pole}
	L_{reg}\otimes\left[\begin{array}{cc}
		Y & I\\ I & X \end{array} \right]+M_{reg}\otimes(\psi+\psi^T)<0,
\end{equation}
where $ \otimes $ denotes Kronecker product and
\begin{equation}\label{psi}
	\psi = \left[\begin{array}{cc}
		AY+B_2\tilde{C} & A+B_2\tilde{D}C_2\\
		\tilde{A} & XA+\tilde{B}C_2
	\end{array} \right].
\end{equation}
The matrices $ L_{reg} $ and $ M_{reg} $ determine the region to which the poles are constrained.



%
%
%
\section*{Conflict of Interest}
The authors declare that they have no conflict of interest.
\bibliographystyle{elsarticle-num}
\bibliography{controlbibfile} 
\end{document}